\documentclass{article}
\usepackage{amsmath,amssymb,amsfonts}
\usepackage{color}
\usepackage{fancyhdr,graphicx}
\usepackage[english]{babel}
\usepackage{hyperref}
\usepackage{natbib}
\usepackage{lmodern}
\usepackage[a4paper,
        left=3.4cm,
        right=3.4cm,
        top=3.4cm,
        bottom=3.9cm,
]
{geometry}
\usepackage{authblk}

\author[1,2,3$\dagger$]{Torsten Heinrich}
\author[1,2]{Juan Sabuco}
\author[1,4,5]{J. Doyne Farmer}

\affil[1]{{Institute for New Economic Thinking at the Oxford Martin School, University of Oxford, Oxford OX1 3UQ, UK}}
\affil[2]{{Oxford Martin School, University of Oxford, Oxford OX1 3BD, UK}}
\affil[3]{{Department for Business Studies and Economics, University of Bremen, 28359 Bremen, Germany}}
\affil[4]{{Mathematical Institute, University of Oxford, Oxford OX2 6GG, UK}}
\affil[5]{{Santa Fe Institute, Santa Fe, NM 87501, USA}}
\affil[$\dagger$]{\textit{torsten.heinrich@oxfordmartin.ox.ac.uk}}

\title{A simulation of the insurance industry:  The problem of risk model homogeneity}

\begin{document}

\maketitle

\begin{abstract}
 We develop an agent-based simulation of the catastrophe insurance and reinsurance industry and use it to study the problem of risk model homogeneity.  The model simulates the balance sheets of insurance firms, who collect premiums from clients in return for ensuring them against intermittent, heavy-tailed risks.  Firms manage their capital and pay dividends to their investors, and use either reinsurance contracts or cat bonds to hedge their tail risk.   The model generates plausible time series of profits and losses and recovers stylized facts, such as the insurance cycle and the emergence of asymmetric firm size distributions.  We use the model to investigate the problem of risk model homogeneity.  Under Solvency II, insurance companies are required to use only certified risk models.   This has led to a situation in which only a few firms provide risk models, creating a systemic fragility to the errors in these models.  We demonstrate that using too few models increases the risk of nonpayment and default while lowering profits for the industry as a whole.  The presence of the reinsurance industry ameliorates the problem but does not remove it.  Our results suggest that it would be valuable for regulators to incentivize model diversity.  The framework we develop here provides a first step toward a simulation model of the insurance industry, which could be used to test policies and strategies for capital management.

\end{abstract}

\section{Introduction}
\label{sect:introduction}

The modern insurance system\footnote{The first insurance contracts emerged much earlier, but the related practices bore little resemblance of the modern insurance system.} has its roots in the establishment of Lloyd's of London in the 1680s, which was named for a coffee house that catered to marine insurance brokers. The first major crisis followed less than a decade later after the Battle of Lagos in 1693. During this battle a fleet of French privateers attacked an Anglo-Dutch merchant fleet causing estimated losses of around 1 million British pounds\footnote{This event is also known as the Smyrna fleet disaster and was national tragedy for England. The value of the English GPD in the year 1693 is estimated to be around 59 million pounds \citep{Campbell09}.} \citep{Leonard13b,Go09,Anderson00}. Risk assessment was inadequate and underestimated several risk factors.\footnote{A large number of merchant ships (around 400) with extraordinarily valuable cargo travelled together and were therefore vulnerable to the same risks. In addition, the measures taken to protect the fleet (22 escorting warships) were inadequate. 
} Worse, it was not only some few underwriters that took the risk of writing policies for this merchant fleet, it was a significant part of the industry. $33$ underwriters went bankrupt. The English parliament considered legislation that would have resulted in a government bailout~\citep{Commons}, but the bill failed \citep{Leonard13b,Leonard13}. 

Today's insurance-reinsurance systems build on centuries of experience.  Modern insurance companies have moved beyond the coffee house and are built on a more solid institutional foundation, and are hopefully more prudent and more competent in assessing risks.   Nonetheless, the example serves to illustrate two points. First, catastrophic damages are difficult to anticipate with any accuracy and unanticipated high losses remain a reality. More recent examples include the Asbestos case, the Piper Alpha disaster, and the 2017 Caribbean hurricane season. Second, a lack of diversity in risk models can lead to problems at a systemic scale.

The insurance industry has made huge progress in its ability to estimate risks.  But there is more to the insurance business than simply estimating individual risks.  Companies need to make a variety of decisions, such as how much total risk to take, how much capital to hold in reserve, and how to set premiums.  Insurance companies compete with each other and so they do not make these risks in isolation.   This can lead to systemic effects that create systemic risks that are not visible to individual firms.

Our goal here is to create a model that makes it possible to study systemic effects at the level of the insurance industry as a whole.  To do this we simulate individual firms and the perils they ensure, but we also simulate how firms set premiums, how they manage their capital, and how these actions affect each other.  Our model is the first to simulate the catastrophe insurance industry at this level.  Here we use the model to address a specific problem, relating to the dangers of consolidating the entire industry under a few risk models under Solvency II, which is an insurance regulation framework adopted by the EU in 2009 and implemented after several delays in 2016, and to explore the role of the reinsurance industry in mitigating risks.  However, we think the possible uses of this model go beyond those we explore here.


The business of the retail insurer is to pool and hedge risks and to hold sufficient capital in sufficiently liquid form to compensate for damages when they happen.  This works very well as long as damages are small and uncorrelated. In this case, providing all moments exist and the distribution of damages is known, the central limit theorem makes it possible to estimate damages accurately. For catastrophes, in contrast, the distributions are not always well understood, the tails are typically heavy, and events are not always independent.

The catastrophe insurance system is subject to several nonlinear effects.   Perils such as earthquakes, hurricanes, flooding, and other natural catastrophes occur rarely, but when they do the damages can be significant. Both the intervals between perils and the damage sizes follow long tailed distributions \citep{Emanuel10,Embrechts99,Christensen02}.  Insurers typically take out reinsurance to cover their tail risk; in present times this typically applies to damages beyond 50 million USD and up to 200 million USD. Each insurance or reinsurance firm typically enters into reinsurance contracts with a wide range of other firms, spreading the risk around. Each firm attempts to estimate risks using models that are typically provided by third parties. Modern catastrophe risk models are sophisticated but are inevitably inaccurate, due to the difficulty of fitting heavy tailed, non-stationary distributions. Climate change, for instance, affects the frequency and severity of windstorms, while changing settlement patterns affect damage sizes \citep{Grinstedetal19}.

\begin{figure}[tbhp]
  \centering
  \includegraphics[width=1\textwidth]{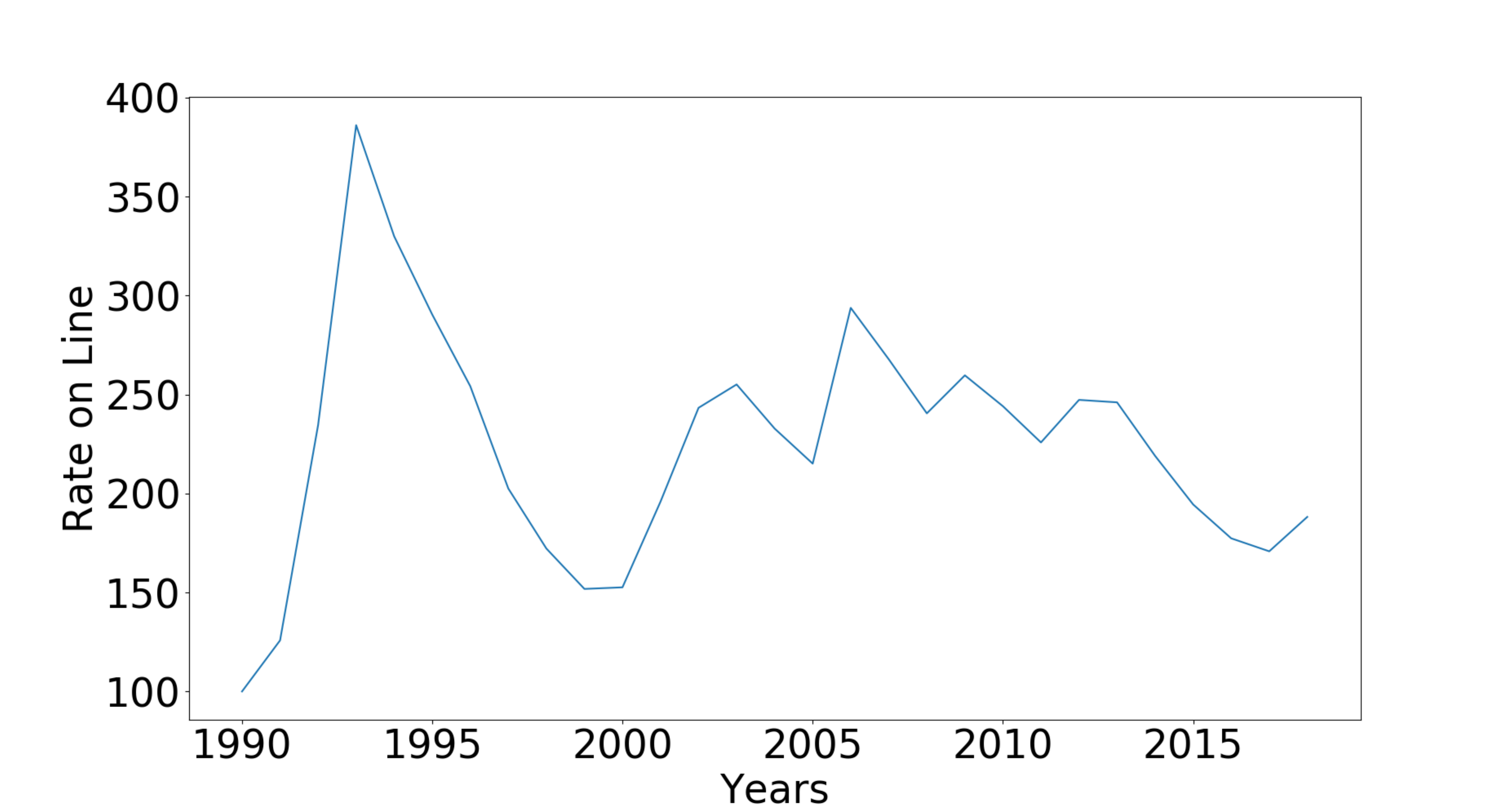}
  \caption{{\it The insurance cycle} is shown here in terms of the global reinsurance Rate-on-Line (ROL) index, which is plotted as a function of time.  The ROL is the ratio of the average premium paid to recoverable losses (shown here in percent). The \textit{ROL index} is computed from the ROLs of contracts and renewals in a number of markets across the globe, and provides a measure of the quality of the market for the industry as a whole.  The striking feature is that the ROL varies across a large range, from $100\%$ to almost $400\%$, and it is highly correlated from year to year. 
 The data is from {\it Guy Carpenter}, who is a reinsurance broker; raw data and details of the computation are not public. }
  \label{fig:global_rate_on_line}
\end{figure}

The insurance industry has historically tended to oscillate between strongly competitive periods, called {\it soft markets}, and less competitive periods, called {\it hard markets}.  This oscillation is usually referred to as the \textit{insurance cycle} or \textit{underwriting cycle}.   As shown in Figure~\ref{fig:global_rate_on_line}, the insurance cycle is irregular in both period and amplitude.  During soft markets competition increases, premiums are down, insurers are more willing to underwrite new business, underwriting criteria are relaxed and the capacity of the industry increases.  Conversely, during hard markets premiums are high, competition is low, underwriting criteria are tight, and the capacity of the industry contracts.  For catastrophe insurance the cycle bears some relationship to the incidence of catastrophes, but it does not track this as closely as one might expect, and depends also on other factors, such as available capital.   The insurance cycle occurs even in sectors such as life insurance and casualty insurance, that do not suffer large fluctuations in claims.  There is no consensus about the cause of the insurance cycle (though see \citet{Zhou13,Owadallyetal18} for an example of a model for cycles in other industries).  During recent years the amplitude of the insurance cycle has been reduced, which some attribute to an increased influx of capital.  
In this paper we build an agent-based simulation model that includes insurance customers, insurance firms, reinsurance firms, catastrophe bonds, and shareholders. 
Insurers and reinsurers choose from a selection of one or more risk models, whose accuracy and homogeneity we control. Based on these, insurers and reinsurers make decisions to underwrite contracts or not. Catastrophe events happen at random, lead to claims, and potentially to bankruptcies, if an insurer or reinsurer is unable to pay. 
The model reproduces a wide range of known stylized facts about the insurance-reinsurance sector, generating stationary dynamics for the insurance-reinsurance system,\footnote{The realizations vary substantially with random events, but the ensemble average and distribution becomes static after a transient period. Further, that state represents a viable insurance-reinsurance system. In almost all time periods in almost all realizations the system is able to satisfy a large share of the demand.} a long-tailed firm size distribution for both insurance and reinsurance firms and a realistic insurance cycle.   

We apply our model to address a controversial aspect of the European regulatory framework for the insurance industry, which is called Solvency II.  This framework was adopted by the EU in 2009 and implemented after several delays in 2016. It includes requirements for capital, liquidity, and transparency on the part of the insurance companies.  In addition it also includes standards for risk models, which must be certified.  While the resulting quality control likely has substantial beneficial effects, there are concerns that the perspective of Solvency II is too strongly microprudential, and may ignore possible negative effects at the systemic level \cite{Gatzert/Wesker12}.  

There are concerns about systemic fragility that may be caused by using a small number of risk models.  In part because of the strick requirements of Solvency II, insurance firms have been driven to outsource the problem of risk modeling.  The provision of risk models has come to be dominated by three major competitors, RMS (Risk Management Solutions), AIR (Applied Insurance Research) and EQECAT.  According to a survey carried out by the Bermuda Monetary Authority, these three vendors ``practically control the entire catastrophe modelling sector" \citep{BMA/Share}.   In Table \ref{tab:cat_models_per} we show the relative overall usage of risk models by insurance firms from 2011 - 2017.  A few overall trends are apparent:  Many firms only use a single risk model.  The dominant provider, RMS, is used exclusively by $41\%$ of the firms in 2017, rising from $33\%$ in 2011.  The second largest provider, AIR, is used exclusively by $19\%$ of firms, and the third largest provider, EQECAT, is not used exclusively by any firms.  The remaining firms use a mixture of models by RMS and AIR.  In 2001 $21\%$ of the firms used a mixture of models by all three providers; there are now no firms that do this.  

\begin{table}[tbhp]
\centering
\textbf{ Vendor Models Usage (In percent of respondents)}
\begin{tabular}{|p{4cm} | c | c | c | c | c | c | c |}
\hline\hline
                         & \textbf{2011}  & \textbf{2012}  & \textbf{2013} & \textbf{2014} & \textbf{2015} & \textbf{2016} & \textbf{2017} \\
\hline\hline
AIR                     &  6.1 &  8.8 & 11.4 & 16.7 &  9.1 & 12.5 & 18.9 \\
EQECAT                  &  0.0 &  0.0 &  0.0 &  0.0 &  0.0 &  0.0 &  0.0 \\
RMS                     & 33.3 & 26.5 & 28.6 & 30.6 & 39.4 & 40.6 & 40.5 \\
AIR and RMS             & 36.4 & 44.1 & 45.7 & 38.9 & 45.5 & 43.8 & 40.5 \\
AIR and EQECAT          &  0.0 &  0.0 &  0.0 &  0.0 &  0.0 &  0.0 &  0.0 \\
EQECAT and RMS          &  3.0 &  2.9 &  0.0 &  0.0 &  0.0 &  0.0 &  0.0 \\
AIR, EQECAT, and RMS    & 21.2 & 17.6 & 14.3 & 13.9 &  6.1 &  3.1 &  0.0 \\\hline\hline 
One risk model          & 39.4 & 35.3 & 40.0 & 47.3 & 48.5 & 53.1 & 59.4 \\
Two risk models         & 39.4 & 47.0 & 45.7 & 38.9 & 45.5 & 43.8 & 40.5 \\
Three risk models       & 21.2 & 17.6 & 14.3 & 13.9 &  6.1 &  3.1 &  0.0 \\\hline\hline  
\end{tabular}
\caption{{\it Market shares of catastrophe risk modelers in Bermuda}, presenting data from a survey conducted by the Bermuda Monetary Authority \citep{BMA/Share,BMA/Share18} on models used by insurers and reinsurers. The table shows low and declining risk model diversity. 
}\label{tab:cat_models_per}
\end{table}

There are good reasons why using more than one risk model is desirable.  Risk models are inevitably inaccurate, and using more than one makes it possible to average out inaccuracies and improve forecasts.  This is a consequence of the general desirability of forecast combination\footnote{
See  \cite{Bates/Granger69,Newbold/Granger74,Deutsch/Granger/Teraesvirta94,Hong/Page04}.
}.
Given a set of models, each of which makes useful forecasts that are not perfectly correlated, forecasts can be improved by taking weighted averages of the forecasts of each model.  This provides an incentive for individual firms to use more than one model.  From a systemic point of view, it is desirable for different insurance firms to use different models.  This is because of one firm goes bankrupt due to errors in its model, firms with different models (and different errors) may survive as a result of their diversity.  Thus even if one provider's risk models are superior to those of other providers, it may still be desirable for the industry as a whole if firms use a diversity of models from different providers.

We study this problem here and demonstrate that the diversity of available risk models has a strong effect on both the distributions of the sizes of bankruptcy cascades and the integrity and capacity of the insurance system. We further investigate the effect of reinsurance by running counterfactual simulations without reinsurance, as well as the sensitivity with respect to parameters of the model. We find that reinsurance can mitigate this problem to a certain extent - without reinsurance, the impact of risk model homogeneity is even stronger. However, reinsurance also introduces a second contagion channel (counterparty exposure from reinsurance): For all settings with more than one risk model (i.e., with no absolute risk model homogeneity), we find that large bankruptcy cascades affecting more than 10\% of the firms are about 20\% less frequent\footnote{This is quantified using a large ensemble of 400 replications of 4000 months (333 years) each.} as discussed in more detail in Section~\ref{sect:results:reinsurance}.

The paper is organized as follows: Section \ref{sect:literature} gives an overview of previous work, Section \ref{sect:model} introduces the model. Results are discussed in Section \ref{sect:results}. Section \ref{sect:conclusion} concludes.

\section{Literature review}
\label{sect:literature}

We will next discuss the state of the art of models of the insurance sector (Section~\ref{sect:literature:modeling}), that our model could potentially build upon. Sections~\ref{sect:literature:insurancecycle} and \ref{sect:literature:systemicrisk} review previous findings on two applications of our model, the modeling of the insurance cycle (Section~\ref{sect:literature:insurancecycle}) and the investigation of systemic risk in insurance systems (Section~\ref{sect:literature:systemicrisk}).
We discuss empirical findings that may be used for calibration in Section~\ref{sect:literature:empirics}. How these stylized facts are reflected in the model design is discussed in more detail in Section~\ref{sect:model}. 

\subsection{Modeling the insurance sector}
\label{sect:literature:modeling}

Very few agent-based models of the insurance sector have been developed so far, although there are three notable exceptions: The London flood insurance model by \cite{Dubbelboer17}, the model of premium price setting in non-catastrophe retail insurance by Zhou \citeyearpar{Zhou13}, further developed in \cite{Owadallyetal18}, and the agent-based model extension to Maynard's study on catastrophe risk \citep{Maynard16}.

Non-agent-based analytical contributions often take an equilibrium approach based on game-theory and common assumptions of frictionless markets and rational decision-making. While this can offer some basic guidance on modeling specific elements of insurance markets, their value for system-level analyses and for predictions is limited due to strong assumptions. An example is the hypothesis of the ``square-root rule of reinsurance'' \citep{Powers/Shubik06}, that derives the optimal relation of the number of reinsurers to that of insurers as following a square-root function of the size of the system. While the empirical relationship is indeed sub-linear, studies \citep{Venezian05,Duetal15} cannot confirm the exact square-root nature. Other examples include Plantin's \citeyearpar{Plantin06} model of the reinsurance market, which aims to prove that reinsurance is necessary for a functioning insurance sector and profitable as a business model under normal conditions. It proceeds to assume that under rationality assumptions, some insurance firms will become reinsurers if the reinsurance sector is not sufficiently large, while making no comment about the dynamics of and possible friction in this process. The limitations of models of this type are well-known in the literature (see, e.g., \citealp{Powers/Ren03}). These limitations can potentially be overcome by agent-based models.

\citet{Zhou13} and \citet{Owadallyetal18} consider pricing in non-life insurance. Risk modeling, systemic effects, and catastrophes are side-aspects in this model. Data used to validate the model is taken from the motor insurance sector of the UK, where catastrophic damages at system-scale are unlikely. The study considers various pricing strategies and is able to recover a realistic insurance cycle with direct local interactions (as opposed to a centralized market) being a major factor. They conclude that the insurance cycle cannot be solely driven by repeated catastrophic shocks.

\citet{Maynard16} investigates whether the use of scientific models can improve insurance pricing. An agent-based approach is used to evaluate how useful those forecasts are in systems with competing insurance firms. To remove interference from other effects, the number of companies is limited to two and the forecasting strategies are fixed, which makes it possible to investigate survival time and commercial success in a controlled setting.

\citet{Dubbelboer17} explores the dynamical evolution of flood risk and vulnerability in London. This agent-based model is used to study the vulnerability of homeowners to surface water flooding, a major source of catastrophe risks in the United Kingdom. The model focuses on the role of flood insurance, specially in the public-private partnership between the government and insurers in the UK, and the UK's flood re-insurance scheme {\it Flood Re}, which has been introduced as a temporary measure for 25 years starting in 2014 to support the development of a functioning flood re-insurance sector in the country.

In contrast to these approaches, we aim to construct a more comprehensive, generic, and flexible agent-based model of the insurance sector, as introduced in Section~\ref{sect:model}.


\subsection{The insurance cycle}
\label{sect:literature:insurancecycle}


There is no consensus in the literature on the causes of the insurance cycle. One major literature tradition believes that natural disasters and large catastrophes are the main driving force (cf. \citealp{Lamm-Tennant/Weiss97}). Such events are believed to trigger the transition from a soft market to a hard market. After a catastrophe, the insurance industry receives a large amount of claims that deplete the capital of most insurers while driving those that are less capitalized out of business. The ones surviving reconsider their underwriting criteria, are more reluctant to take risks, and premiums start rising as a consequence. Mergers and acquisitions activity also increases during a hard market, especially after a catastrophe when the claims start depleting the capital of the industry. This may already start in the immediate aftermath of the event before any claims are filed, as market participants anticipate substantial damages. The mergers and acquisitions activity also contributes to the reduction of capital in the industry and the increase of prices, since the surviving firms have to absorb the losses of the firms that go out of business and possibly also since they enjoy more market power. In reality, very few firms in the sector file for bankruptcy since the ones in financial difficulties are absorbed by the better capitalized ones due to the value of their customers, branding, insurance information and human capital.

This literature tradition is exemplified by \citet{Lamm-Tennant/Weiss97}, who aim to identify the insurance cycle empirically by fitting an AR(2) process and to explain its existence and period by running regressions with incidence of catastrophe events and various other explaining variables. They find that catastrophe events are significant while many other variables are not. A drawback of this analysis is that the time series considered are only 20 years long, though they include data for a number of countries. 

Other contributions, most notably in this context the ABM analysis by \citet{Zhou13} and \citet{Owadallyetal18}, contradict this explanation, as they are able to model the emergence of an insurance cycle from price effects without any catastrophe events.

\subsection{Systemic risk in insurance}
\label{sect:literature:systemicrisk}


The problem of systemic risk in insurance came into focus after the reinsurer AIG became illiquid and had to be bailed out by the US government during the financial crisis in 2008. \citet{Park/Xie14} conduct a stress test and find that the systemic damage resulting from one big reinsurer defaulting in the US market would be very limited. \citet{Cummins/Weiss14} are more cautious; they point out that there is significant counterparty exposure within the reinsurance market through retrocession,\footnote{Reinsurance of reinsurance companies is called retrocession.} and that this is exactly what led to the historic near-meltdown of the insurance system in the LMX spiral.\footnote{Reinsurers at the Lloyd's of London were present in multiple layers of retrocession without realizing it. When the disaster at the Piper Alpha oil rig in 1988 caused unanticipated high losses, these retrocession layers were triggered, hitting already cash-strapped firms again and at the same time denying lower layers a speedy recovery of claim payments. The retrocession branch and the insurance business as a whole have become more prudent in this regard \citep{Cummins/Weiss14}.} 

\citet{Cummins/Weiss14} further note possible challenges from other aspects of the system, such as the size distribution of the firms and interconnections with asset markets. This hints at other contagion channels of systemic risk besides counterparty exposure. As in banking, portfolio similarity may be a serious compounding factor in the case of sell-offs, and the interaction of multiple contagion channels may aggravate systemic risk disproportionately \citep{Cacciolietal15}.

Solvency II has been hailed for its capacity to decrease capital and liquidity risk \citep{Ronkainenetal07,Gatzert/Wesker12}. Even authors with a macroprudential focus \citep{Gatzert/Wesker12,Kessler14} judge systemic risk in modern insurance systems with up-to-date regulation (and explicitly Solvency II) as unlikely. However, their assessment is limited to contagion channels present in banking systems. In this regard, insurance firms, which are not highly leveraged, appear quite safe. However, \cite{Elling/Pankoke14} voice concerns regarding a potential pro-cyclicality of the Solvency II regulation framework.

A final contagion channel, the one investigated in Section~\ref{sect:results:risk-model-diversity}, may be caused by risk model homogeneity. This has been mentioned in passing remarks \citep{Petratosetal17,Tsanakas/Cabantous18}, but the present study is to our knowledge the first to investigate this issue in a systematic way.

\subsection{Empirical findings}
\label{sect:literature:empirics}

Empirical research relevant to agent-based model development in the field of catastrophe insurance includes studies on the insurance-reinsurance system by \citet{Froot01} and \citet{Garven/Lamm-Tennant03} as well as Boyer and Dupont-Courtade's \citeyearpar{Boyer/Dupont-Courtade15} analysis of reinsurance programmes.\footnote{Insurers tend to slice all the risks in a peril region or the whole firm in various layers of reinsurance by damage size. Each layer is ceded in different contracts and likely to different reinsurance firms.} All three papers use proprietary data sets. 

Traditional wisdom holds that the insurance cycle is mainly driven by the steady stream of catastrophe events. \citet{Froot01} reports extensive data on reinsurance pricing and shows that prices (in terms of the relation of premium to expected loss) have decreased in the second half of the 1990s, i.e. in recent years before his paper was published. He states that the absence of large catastrophic events during this time frame is the main reason for this decrease of premiums, but also mentions the alternative interpretation of an insurance cycle driven by a mechanism different from catastrophe events. \citet{Garven/Lamm-Tennant03} find, perhaps unsurprisingly, that demand for reinsurance decreases with the firm size of the insurance firm buying reinsurance (the \textit{ceding insurance firm}) and its concentration in line-of-business and location and increases with leverage of the ceding firm and with the tail weight (thus, risk) of it's written insurance. \citet{Boyer/Dupont-Courtade15} discuss the layered structure of reinsurance programmes. Data reported in the paper shows that treaties with one to five layers are common\footnote{They do, however, report a slightly decreasing frequency with the number of layers in their data set: 277 one-layered treaties, 201 two-, 235 three-, 129 four-, 79 five-layered treaties.}, but much larger treaties with up to eleven layers occur. A temporary decrease of the number of contracts with the financial crisis in 2008 is evident in their data. They report that parameters of the contracts (premium of the accepted bid, dispersion of the received bids) vary widely across the lines of business. Higher layers tend to be cheaper in terms of rate-on-line (defined as premium divided by limit), as losses affecting these layers are less likely albeit potentially heavy. 

The amount of capital used to support reinsurance worldwide has been growing quickly. Most of the growth continues to come from reinsurer and insurer profits and investments, but a substantial amount of capital has recently been injected from sources that did not exist 20 years ago. While these alternative capital sources have almost no impact on the typical policyholder, they have significantly affected the way reinsurance is currently written worldwide. Catastrophe bonds (also known as CAT bonds) \citep{Cummins08} are securities that allow the transfer of risks from insurers and reinsurers to institutional investors like hedge funds, mutual funds and pension funds. CAT bonds are attractive to these investors since they have a relatively low correlation with the rest of the financial market and allow the investors to achieve higher diversification. The CAT bond market has been growing steadily over the last 20 years and may have contributed to dampen the insurance cycle. The analysis of CAT bonds by \citet{Lane/Mahul08} shows that the equivalent measure for CAT bonds to the premium, {\it spread at issue} over {\it LIBOR}, is explained quite well by a simple linear model (spread at issue as a function of {\it expected loss}) although there are other minor influences,\footnote{Data from Lane Financial LLC. seen by the authors of the present paper confirm this.} which make it possible to model the pricing of these instruments in a rather simple way.  


\begin{figure}[tbhp]
  \centering
  \includegraphics[width=.7\textwidth]{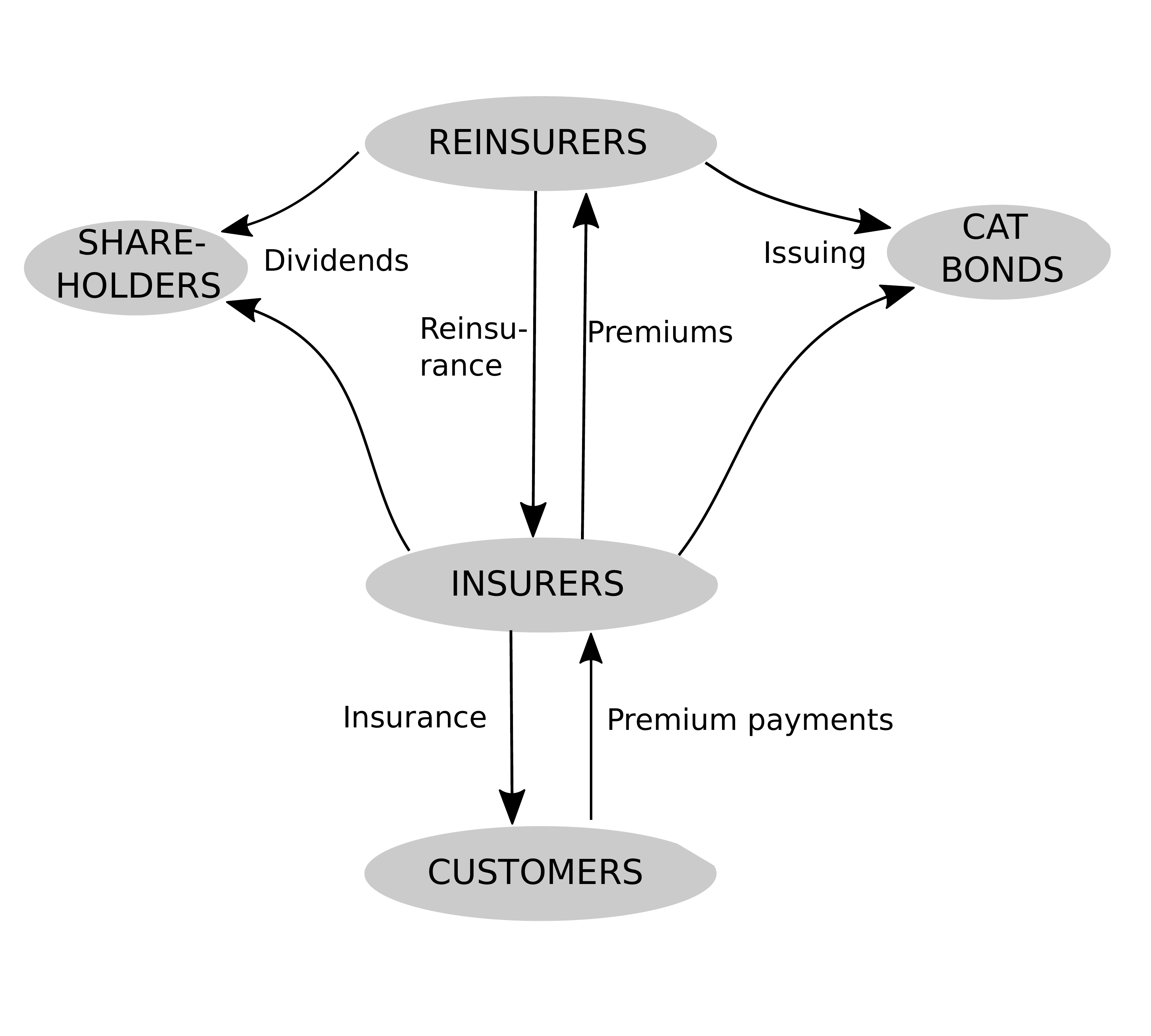}
  \caption{Agents and interaction structure of the agent-based model.}
  \label{fig:model-description-scheme}
\end{figure}

\begin{figure}[tbhp]
  \centering
  \includegraphics[width=.7\textwidth]{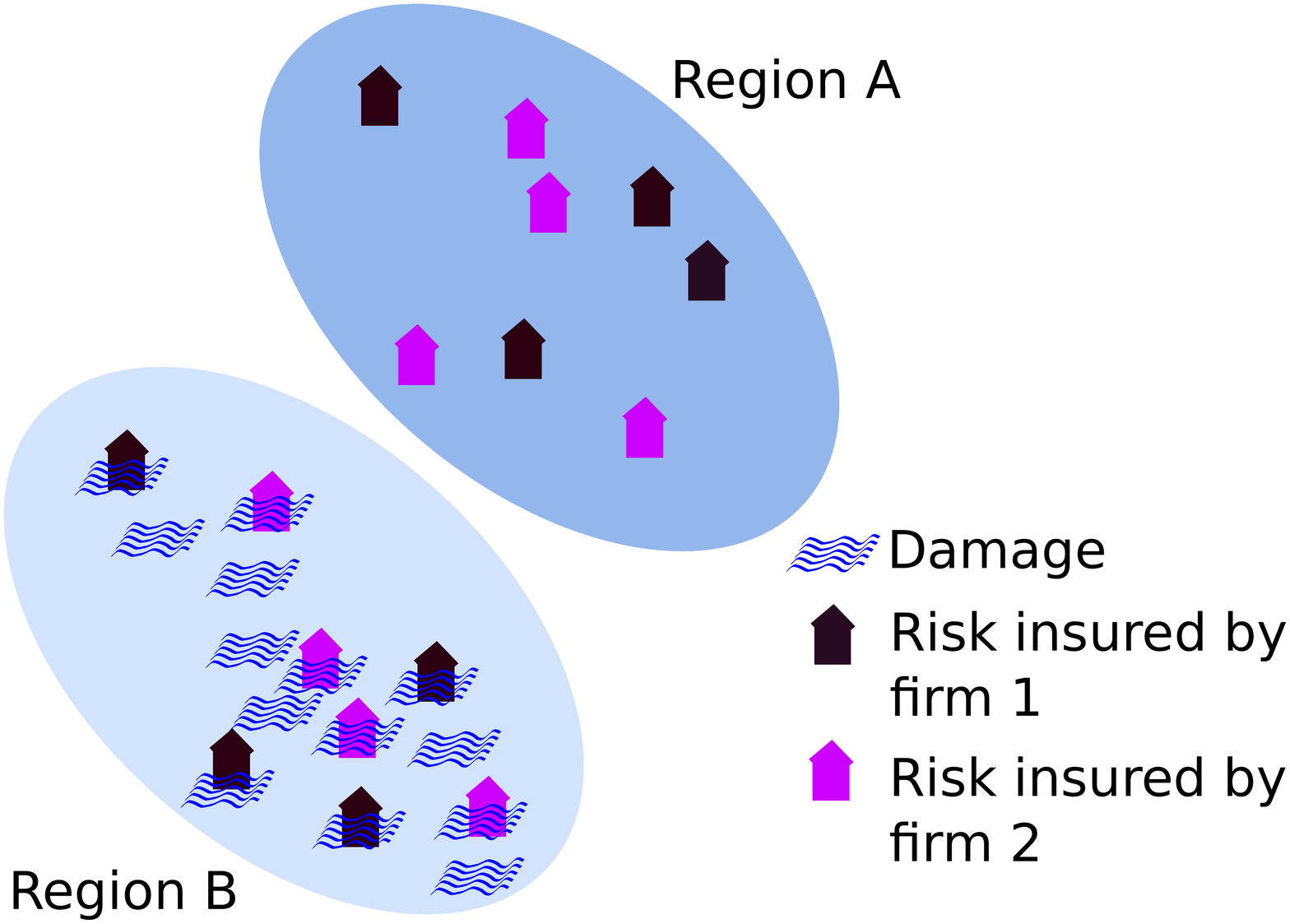}
  \caption{{\it Peril regions and firms.}  A given insurance firm may issue insurance in multiple peril regions.  When a damage occurs in a given peril region, we assume it affects all the policies issued in that region, leading to multiple claims with all the firms involved, but it does not affect policies issued in other peril regions. Claims in each peril region therefore occur in clusters.}
  \label{fig:peril-regions}
\end{figure}

\section{Model description}
\label{sect:model}

Based on stylized facts from the literature in the previous section, we develop an agent-based model of insurance-reinsurance systems. To make it possible to study both systemic aspects and characteristics of individual elements, we choose a modular design, so that agents of different types can be switched on and off individually.  We discuss a range of relevant application in Section~\ref{sect:results}.


\subsection{Agents}

The model, illustrated in Figure~\ref{fig:model-description-scheme}, includes five types of agents: insurance customers, insurers, reinsurers, shareholders, and catastrophe bonds. Customers buy insurance coverage and pay premiums. Insurance firms may obtain reinsurance from either traditional reinsurance companies or catastrophe bonds. Insurance and reinsurance contracts oblige the customer (or the insurer obtaining reinsurance) to make regular premium payments, but entitle them to claim reimbursements for covered damages under certain conditions. 

Insurance and reinsurance firms (discussed in more detail below) are the core of the model. Most of the decision making capacity in the model lies with them. They consult risk models to support their decision making. They further pay dividends to shareholders.


\subsection{Customer side}

\subsubsection*{Insurance customers (households)}

Customers are modeled in a very simple way. They own insurable risks which they attempt to insure. They approach one insurer per time step and accept the current market premium if the insurer offers to underwrite the contract. The value of the insurable risks is normalized to $1$ monetary unit each and the total number of insurable risks is fixed. The risks are not destroyed but are assumed repaired to their previous value after each damage incident.\footnote{Identical or similar values for all risks is fairly realistic for property insurance, which is a large part of of the catastrophe insurance business. Other types of insurance (ships, airplanes, satellites, etc.) are subject to size effects on the part of the individual risk and may therefore show a different behavior, but may have similar characteristics to the excess-of-loss reinsurance business included in our model, which also involves individual contracts of large value.} 

\subsubsection*{Perils and peril regions}

It is convenient to distinguish catastrophic and non-catastrophic perils. Catastrophic perils are the ones affecting most of the risks of a particular peril region, e.g. resulting from a hurricane in Florida, an earthquake in Japan or a flood in Southern England. While perils are rare, they can lead to heavy losses and are thus a primary reason for reinsurance. Non-catastrophic perils, on the other hand, typically affect only individual risks and are are more frequent and uncorrelated in time. Examples of this type of perils are car accidents, residential fires or retail burglaries. 

In our model we only consider catastrophic perils, assuming that the effect of the non-catastrophic ones is minor, sometimes covered by deductibles, and subject to averaging out across many risks as a result of the central limit theorem. Catastrophic perils are modelled as follows:
\begin{itemize}
  \item Catastrophe event times are determined by a Poisson process, i.e. event separation times are distributed exponentially with parameter $\lambda$. For more details see Appendix~\ref{app:time_dist}.
  \item Total damage follows a power-law with exponent $\sigma$ that is truncated at total exposure (since insurance payouts cannot be higher than the amount insured). For more details about this see Appendix~\ref{app:global_loss}.
  \item Total damage is assigned to individual risks following a beta distribution calibrated to add up to the total damage. For more details about this see Appendix~\ref{app:individual_loss}.
\end{itemize}
In the model each insurable risk belongs to one of $n$ peril regions, see Figure~\ref{fig:peril-regions}. For simplicity we assume that all the risks of the respective peril region are affected by a catastrophic peril. In this study we typically consider $n=4$.

\subsection{Insurer side}

\subsubsection*{Firms, capital, entry, exit}

The number of firms in the model at time $t$ is $f_{t}=i_{t}+r_{t}$, of which $i_{t}$ are insurance firms and $r_{t}$ are reinsurance firms. The number of firms is dynamic and endogenous with initial values $i_{0}$ and $r_{0}$. 

Market entry is stochastic with constant entry probabilities for insurers ($\eta_i$) and reinsurers ($\eta_r$). New insurance firms have a given initial capital $\overline{k}_i$ and new reinsurance firms have initial capital $\overline{k}_r$.  These are both constants, chosen so that   $\overline{k}_r$ is substantially larger than  $\overline{k}_i$.

Market exit occurs with bankruptcy or when insurers or reinsurers are unable to find enough business to employ at least a minimum share $\gamma$ of the cash that they hold for $\tau$ time periods. (We calibrate the model so that one time period is roughly a month).  Since the return on capital would be extremely low in that case insurers and reinsurers prefer to leave the market or focus on other lines of business. We typically set the parameters to $\gamma_i=0.6$, $\tau_i=24$ for insurance firms and to $\gamma_r=0.4$, $\tau_r=48$ for reinsurance firms. That is insurance firms exit if they employ less than 60\% of their capital for 24 months, reinsurance firms when they employ less then 40\% of their capital for 48 months. 

Firms obtain income from premium payments and interest on capital $k_{j,t}$ (of firm $j$ at time $t$) at interest rate $\xi$. Firms also cover claims, and may attempt to increase capacity by either obtaining reinsurance or issuing CAT bonds. They pay dividends at a rate $\varrho$ of profits as shown in appendix~\ref{app:dividends}. Dividends are only paid when there are profits. Firms decide whether or not to underwrite a contract based on whether their capital $k_{j,t}$ can cover the combined \textit{value-at-risk} (VaR) of the new and existing contracts in the peril region with an additional margin of safety corresponding to a multiplicative factor $\mu$. They additionally try to maintain a diversified portfolio with approximately equal values at risk across all $n$ peril regions.  

Policyholders, shareholders, catastrophe bonds, and institutional investors that would buy catastrophe bonds (such as pension funds and mutual funds ) are not represented as sophisticated agents in this model. Shareholders receive dividend payments. 
Institutional investors buy catastrophe bonds at a time-dependent price that follows the premium price. They do not otherwise reinvest or have any impact on the companies' policies. 

CAT bonds pay claims as long as they are liquid, and are dissolved at bankruptcy or otherwise at the scheduled end of life (at which point the remaining capital is paid out to the owners). The modular setup of the ABM allows us to run replications with and without reinsurance and CAT bonds. 

\begin{table}[tbh]
\begin{tabular}{p{2.cm}|p{1.6cm}|p{0.7cm}|p{0.7cm}|p{0.7cm} p{0.2cm} p{0.2cm} p{1.1cm}|p{1.1cm}|p{1.1cm}|p{1.1cm}}
\multicolumn{7}{c}{\textbf{ Risk model properties}} &
\multicolumn{4}{c}{\textbf{Risk model usage by setting}}\\\hline\hline
             & {\bf Peril\hspace{0.2cm} {region} A} & {\bf PR B} & {\bf PR C} & {\bf PR D} &&& {\bf Setting 1} & {\bf Setting 2} & {\bf Setting 3} & {\bf Setting 4} \\\hline
Risk model 1 & {\bf U} & + & + & + &&& 100\% & 50\% & 33.3\% & 25\% \\
Risk model 2 & + & {\bf U} & + & + &&&  0\%  & 50\% & 33.3\% & 25\% \\
Risk model 3 & + & + & {\bf U} & + &&&  0\%  &  0\% & 33.3\% & 25\% \\
Risk model 4 & + & + & + & {\bf U} &&&  0\%  &  0\% & 0\%    & 25\% \\\hline\hline
\end{tabular}
    \caption{Risk model diversity (underestimated (U) and overestimated (+) peril regions) and risk model usage by risk model diversity setting (right).}
    \label{tab:riskmodels}
\end{table}

\subsubsection*{Risk models}

\paragraph{VaR.} Each insurance and reinsurance firm employs only one risk model. It uses this risk model to evaluate whether it can underwrite more risks (or not) at any given time. We assume that risk models are imperfect in order to allow investigation of effects of risk model homogeneity and diversity. 

There is empirical evidence that risk models are inaccurate. In some peril regions they tend to underestimate risk while in others they overestimate it. In our model risk models are inaccurate in a controlled way: they are calibrated to underestimate risks in exactly one of the $n$ peril regions and to overestimate the risks in all other peril regions by a given factor $\zeta$ (see Table~\ref{tab:riskmodels}). Since the $n$ peril regions are structurally identical, with about the same number of risks and with risk events governed by the same stochastic processes, this allows up to $n$ different risk models of identical quality.\footnote{To investigate the effects of risk model diversity, it is important that the risk models should be of identical quality in order to avoid interference from effects based on quality differences.}

The risk models use the VaR in order to quantify the risk of the insurers in each one of the peril regions. 
The VaR is a statistic that measures 
the level of financial risk within an insurance or reinsurance firm over a specific time frame. It is employed in some regulation frameworks including Solvency II, where it is used to estimate the Solvency Capital Requirement. Under Solvency II, insurers are required to have $99.5\%$ confidence they could cope with the worst expected losses over a year. That is, they should be able to survive any year-long interval of catastrophes with a recurrence frequency equal to or less than $200$ years. 
The probability that catastrophes generating net losses exceeding the capital of the insurer in any given year is $\alpha =\frac{1}{\mathrm{recurrence\, interval}}=\frac{1}{200}= 0.005$. For a random variable $X$ that would represent the losses of the portfolio of risks of the insurer under study, the VaR with exceedance probability $\alpha \in [0, 1]$ is the $\alpha$-quantile defined as
\begin{equation}\label{vardef}
VaR_{\alpha}(X) = \mathrm{inf}\{x \in \mathbb{R} : P(X > x) \leq \alpha\}. 
\end{equation}
This means that e.g. under Solvency II the capital that the insurer is required to hold can be computed with the $VaR_{0.005}(X)$.     

\paragraph{Computation of the VaR.} 
A firm's capital requirements can be derived from the firm's risk model as a margin of safety factor over the VaR of the entire portfolio: companies should hold capital $k_{j,t}$ such that   
$$k_{j,t} \geq \mu VaR(X_1+X_2+X_3+ ... + X_N),$$  
where the $X_i$ represent all sources of cashflow for the company (including investment returns, credit risk,  insurance losses, premium income, expenses, operational failures etc) and $\mu\geq 1$ is a factor for an additional margin of safety. In other words the firm's whole balance sheet from $t_0$ to $t_0+1$ year must be modeled and capital must be sufficient for the firm to have a positive balance sheet $99.5\%$ of the time as a minimum. Due to catastrophes, this condition can occasionally be violated, e.g., if the company takes a loss such that $k_{j}$ is suddenly and severely reduced. In the present model, the companies will in such cases stop underwriting until enough capital is recovered.

\paragraph{Estimation of the VaR in the simulation.}  
Computing the VaR over the firm's portfolio requires computation of the convolution of the distributions of damages and those of the frequency of catastrophes both over time and in all peril regions while also taking into account reinsurance contracts. Reinsurance contracts essentially remove part of the support of the damage distribution and make them non-continuous.\footnote{For reinsurance firms, reinsurance contracts add parts of other companies' risk distributions between the contract's limit and deductible, also making the resulting distribution non-continuous.} Estimating the non-continuous distribution of cashflows would require a Monte Carlo approach. Since this is necessary for every underwriting decision, it would increase the computation time required for the ABM by orders of magnitude. 

We argue that to study the effects of systemic risk of risk model homogeneity, it is not necessary to compute the $VaR$ combined for all peril regions and over the entire year.\footnote{We aim to assess the effect of risk model diversity. As explained below, we consider risk models that underestimate the risk in exactly one peril region, while overestimating that in others. The model setup for this approach is correct as long as the true VaR is between the underestimating and overestimating values returned by the various risk models. In fact, we can expect our estimate to be much closer to the true VaR.} A good approximation of the dynamics of the insurance sector can be obtained by (1) working with the values at risk due to individual catastrophes in the model and (2) considering the VaR separately by peril region and combining the peril regions with a maximum function. 

(1) The focus on individual catastrophes instead of on 12-month periods transforms the timescale in the results of our simulations, but the type of dynamics and the shape of distributions obtained are the same. Evidently, bankruptcies should be more frequent in our approach since we are only holding capital to survive individual catastrophes with a returning period of $200$ years, but not catastrophe recurrence in 12-month period intervals. 
However, bankruptcy frequency is the only aspect that is affected.\footnote{Changing the interpretation of the model's time steps from months to years would correct this aspect exactly. We opt not to do this, as it entails other problems such as the companies' reaction times, as well as loss of the model's present level of detail.}

(2) Further, computationally expensive convolution of distributions across peril regions can be avoided, since a good approximation can be obtained with the maximum function over the VaRs in individual peril regions. 
To see this, consider two extreme cases. If, on the one hand, the separation times of catastrophes were perfectly correlated between all $n$ peril regions and catastrophes would therefore always coincide, we would have $VaR^c=VaR^1+VaR^2+...+VaR^n$. If, on the other hand, catastrophes would never coincide, we would have $VaR^c=\max(VaR^1,VaR^2, ..., VaR^n)$. The first scenario overestimates the VaR; the second underestimates it, by neglecting the probability of the coincidence of multiple catastrophes. In other words, there is a residual VaR term $VaR^{r}$ to account for this:
$$VaR^c=\max(VaR^1,VaR^2, ..., VaR^n) + VaR^{r}.$$
We choose our parameters such that the probability of such a coincidence happening, 
$$\begin{array}{r l}
P_{coincidence}&=1-\binom{n}{0}(1-P_{peril})^n-\binom{n}{1}P_{peril}(1-P_{peril})^{n-1}\\
               &=1-\binom{n}{0}(e^{-\lambda})^n-\binom{n}{1}e^{-\lambda}(e^{-\lambda})^{n-1},
\end{array}$$
is small. Namely, we choose $\lambda=100/3$, $n=4$, hence $P_{coincidence}\approx0.005$. Consequently, our $VaR^{r}$ is small and we can safely run the model by approximating
$$\widetilde{VaR^c}=\max(VaR^1,VaR^2, ..., VaR^n)$$
\begin{equation}
k_{j,t} \geq \mu \widetilde{VaR^c} = \mu \max(VaR^1,VaR^2, ..., VaR^n).
\end{equation}

\paragraph{Balancing of portfolios based on VaR in the simulation.} In addition, and especially when getting close to the limit $k_{j,t} \approx \mu VaR^i$, firms will prefer to underwrite risks in different peril regions such that the portfolio is approximately balanced, keeping a similar amount of risk in every peril region. More specifically, they underwrite a new contract only if the new the standard deviation of the $VaR^*$ in all peril regions is lower with than without this new contract. That is,
\begin{equation}
std(VaR^{1*},VaR^{2*}, ..., VaR^{n*}) > std(VaR^1,VaR^2, ..., VaR^n), 
\end{equation}
where $VaR^{n*}$ would be the value at risk of every peril region is the new contract is accepted.
If the standard deviation is higher, firms will only be willing to accept a new contract if they are already very balanced enough. In other words, the standard deviation computed with the new $VaR^{n*}$ is small compared to the total cash held by the firm:
\begin{equation}
std(VaR^{1*},VaR^{2*}, ..., VaR^{n*}) < \eta\frac{k}{n}, 
\end{equation}
where $\eta \in [0, 1]$ is a parameter that regulates how balanced a firm wants to be and $n$ is the number of peril regions.



\subsubsection*{Premium prices}

The insurance industry is highly competitive. This justifies the assumption that all agents are price takers. Insurance and reinsurance premiums depend on the total capital $K^{T}_{t}=\sum_{j=1}^{f_t} k_{j,t}$ available in the insurance sector. For the sake of simplicity we assume that insurance premiums oscillate around the fair premium as defined in Appendix~\ref{app:premiums}. When the total capital of the industry increases, the premiums paid by a policyholder decrease, and conversely, they increase when the total capital decreases. To avoid unrealistically high volatility, we set hard upper and lower bounds to the premium. These thresholds are implemented in the model as parameters and can be varied, although we have run most of the simulations with values of $70\%$ of the fair\footnote{By \textit{fair premium} we mean in this context a premium that would on average offset the damages and thus lead to zero profits and zero losses.} premium as lower boundary and $135\%$ of the fair premium as upper boundary.  These boundaries are rarely hit.  

Reinsurance prices are implemented in the same way, using Eq.~\ref{pricingeq} of Appendix~\ref{app:premiums}. The lower and upper limits are the same. The only differences are that reinsurance markets are more sensitive to changes in the reinsurance capital market than insurance markets (see Appendix~\ref{app:premiums} for more details), and that the prices in reinsurance only vary with the total capital available in the reinsurance market, $K^{T}_{t}=\sum_{j=1}^{f_t} z_{j,t}k_{j,t}$ (where $z$ is a vector of length $f_t$ such that element $z_{j,t}$ is 1 for reinsurers and 0 for insurers). The higher sensitivity to capital fluctuations of reinsurance markets is captured by using a steeper slope in Eq.~\ref{pricingeq} in the reinsurance case. The capital in the reinsurance market is usually an order of magnitude below the capital available in the insurance market, which is a feature that can be found in reality and is reproduced by the model in the average steady-state values of the capital after careful calibration. 


For the sake of simplicity premiums are the same for all peril regions. This is a base case that allows us to design risk models of identical quality. 

\subsection{Contracts}

\subsubsection*{Insurance and traditional reinsurance contracts}

Insurers provide standard insurance contracts lasting $12$ iterations (months). At the end of a contract the parties try to renew the contract, which leads to a high retention rate.

Insurers may obtain excess-of-loss reinsurance\footnote{The ABM also allows the possibility of using only proportional reinsurance, although we this is not explored in the present study.} for any given peril region. The standard reinsurance contract lasts $12$ iterations (months). The insurer proposes a deductible for the reinsurance contract; the reinsurer will evaluate whether or not to underwrite the contract. Each reinsurance contract has a deductible (i.e. the maximum amount of damages the insurer has to pay before the contract kicks in). In our model, the deductible for each contract is drawn from a uniform distribution between [25\%, 30\%] of the total risk held per peril region by the insurer at the start of the contract.

\subsubsection*{Alternative reinsurance: CAT bonds}

The model also includes a simplified alternative insurance capital market: Both insurers and reinsurers may issue catastrophe bonds (CAT bonds) by peril region. A CAT bond is a risk-linked security that allows institutional investors like mutual funds and pension funds to reinsure insurers and reinsurers. They are structured like a typical bond, where the investors transfer the principal to a third party at the beginning of the contract and they receive coupons (some points over LIBOR) every year for it. If during the validity of the contract no catastrophe occurs the principal is returned to the investor. If a catastrophe occurs the losses of the insurer and reinsurer are covered with the principal until it is exhausted. CAT bonds are  attractive since they are uncorrelated with the other securities available in the financial market. Since institutional investors are risk averse only the high layers of the reinsurance programs with a low probability of loss are covered by CAT bonds.

If insurers  cannot get reinsurance coverage over five or more iterations then they issue a CAT bond. The premium of CAT bonds is a few points over the reinsurance premium of the traditional reinsurance capital market.

\subsection{Model setup and design choices}

\subsubsection*{Settings}

Risk model homogeneity and diversity can be studied by comparing settings with different numbers of risk models used by different firms. In a one risk model case, all firms use the same (imperfect) risk model. In a two risk model case, firms are divided between two equally imperfect risk models, etc. 

\subsubsection*{Experimental design}

We compare $n$ settings with different numbers of risk models $\nu=1, 2, ..., n$. The up to $\nu=n$ different risk models are of identical accuracy and distinguished by underestimating risks in different peril regions. As a consequence, we need to model $n$ different peril regions; we retain this number of peril regions in all settings including the ones with $\nu<n$ different risk models in order to allow for a more direct comparison.

Simulations are run as ensembles of $M$ replications for each of the $n$ settings considered. In every replication we run the model with identical parameters, changing only the original random seed. We typically set $M=400$ and $n=4$, which means we run $4\times 400=1600$ replications of a simulation just varying the number of risk models ($M=400$ each for $\nu = 1, 2, 3, 4$).

The catastrophes in the model are random, but occur at the same time steps for the different model diversity settings to provide a meaningful comparison. If a catastrophe $x$ of size $D_x$ happens in time step $t_x$ in replication $m_x$ for the one risk model case, then a catastrophe of the same size $D_x$ will hit the simulation in the same time step ($t_x$) in the same replication $m_x$ in the two risk model case, in the three risk model case, and in the four risk model case. This allows us to isolate the effects of risk model diversity as the four different settings are exposed to the same sequences of perils.

We run experiments for various parameter values, e.g. to consider the effects of different margins of safety $\mu$ and the effect of the presence or absence of reinsurance.  

We have designed the model so that after transients die out the behavior is stationary.  This allows us to take long time averages of quantities such as the frequency of bankruptcies.  Quantities such as the number of firms and number of reinsurance firms are set initially, but these change in time in response to the other parameters.  After a long time the initial settings become irrelevant.  To avoid biasing our results with data from the transient stage we remove the first 1200 periods (100 years) of the simulations.

\subsubsection*{Software}

The model was written in Python. The source code is publicly available\footnote{See \texttt{https://github.com/INET-Complexity/isle}}.  It is still under development, e.g., with extensions for validation, calibration, and visualization.

\section{Results}
\label{sect:results}

We now demonstrate applications of the model. We aim to highlight its capabilities and demonstrate insights on the behavior of catastrophe insurance systems that can be gained from it. 

Subsection \ref{sect:results:insurance-cycle} discusses the behavior of the model within single replications. Using the example of the insurance cycle, it will be highlighted that the model is able to reproduce realistic time series.


Subsection \ref{sect:results:risk-model-diversity} compares ensemble simulations with four different risk model diversity settings. The four risk model diversity settings correspond to the one, two, three and four risk models used by the firms in the respective simulation. It serves to demonstrate that the model can be used to investigate systemic risk of model homogeneity in insurance. We show time development patterns in the ensemble simulations (premiums, revenue, numbers of active firms), provide evidence of systematic differences between risk model diversity settings and discuss distributions of firm bankruptcy cascade sizes and of amounts of non-recovered claims.

Subsection \ref{sect:results:reinsurance} investigates the effect of reinsurance by comparing simulations in the base scenario with the normal complement of reinsurance firms (as discussed in previous subsection) on the one hand and counterfactual ones without a reinsurance sector on the other. 

Subsection \ref{sect:results:distributions} discusses emerging distributions of firm sizes that reproduce asymmetric firm size distributions in reality nicely in spite of initially equal firm sizes in the model. 

If not indicated otherwise, the parameter settings are as given in table~\ref{tab:parameters} in appendix~\ref{app:parameters}.

\begin{figure}[tbhp]
  \centering
  \includegraphics[width=1\textwidth]{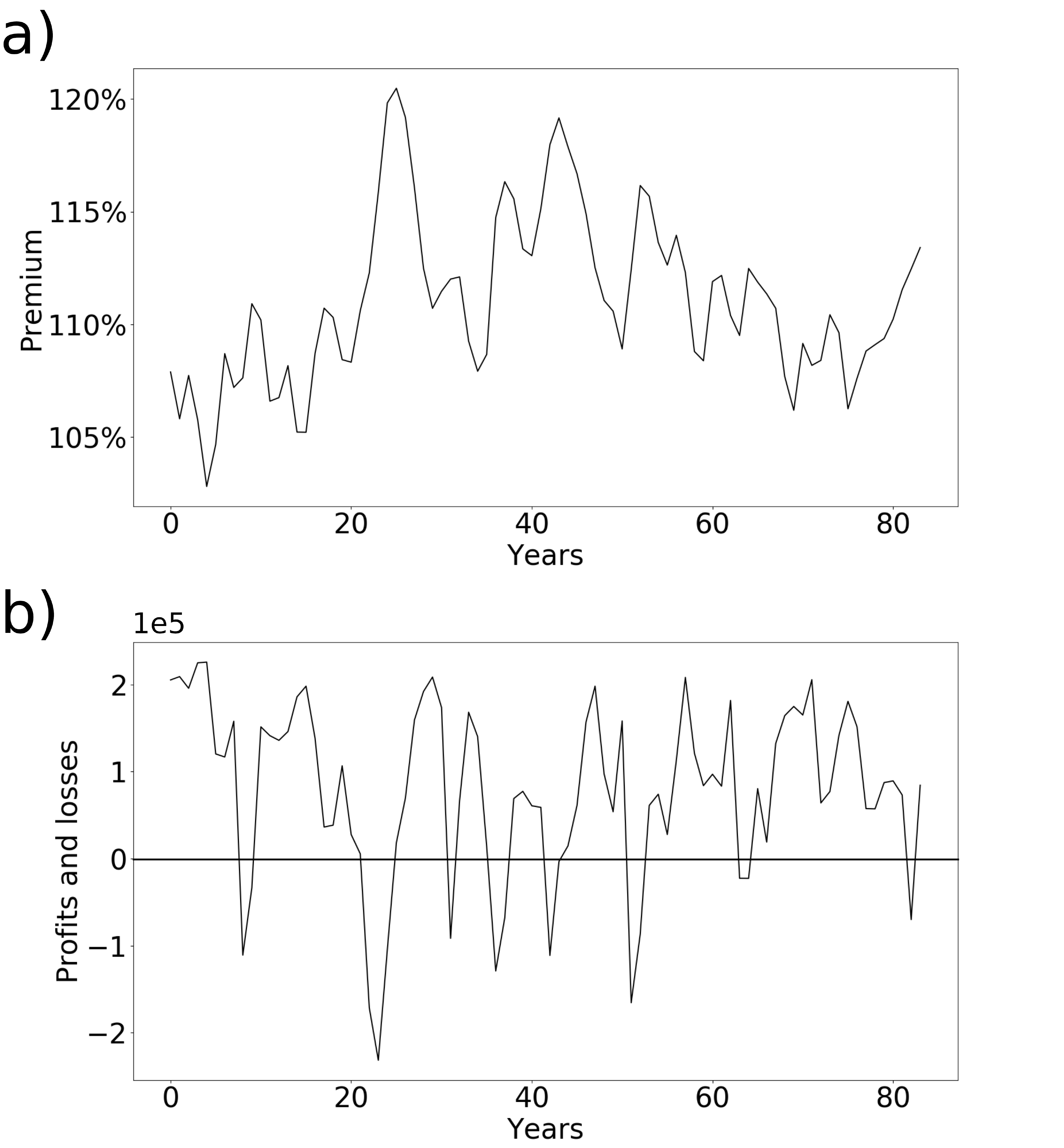}
  \caption{{\it Insurance cycle for the model}.  a)  Time series of the premium in percent of the fair premium (expected loss) in a typical simulation run, where $100\%$ means that premiums are on average equal to claims. 
  b) Time series for the same simulation run of the sum of profits and losses of all insurers  in normalized monetary units.  The insurance cycle emerging in the development of the premium (a) has realistic characteristics and is distinct from (albeit influenced by) the development of profits and losses (b).
  }
  \label{fig:combined}
\end{figure}

\subsection{Reproducing the insurance cycle}
\label{sect:results:insurance-cycle}


\begin{figure}[tbhp]
  \centering
  \includegraphics[width=1\textwidth]{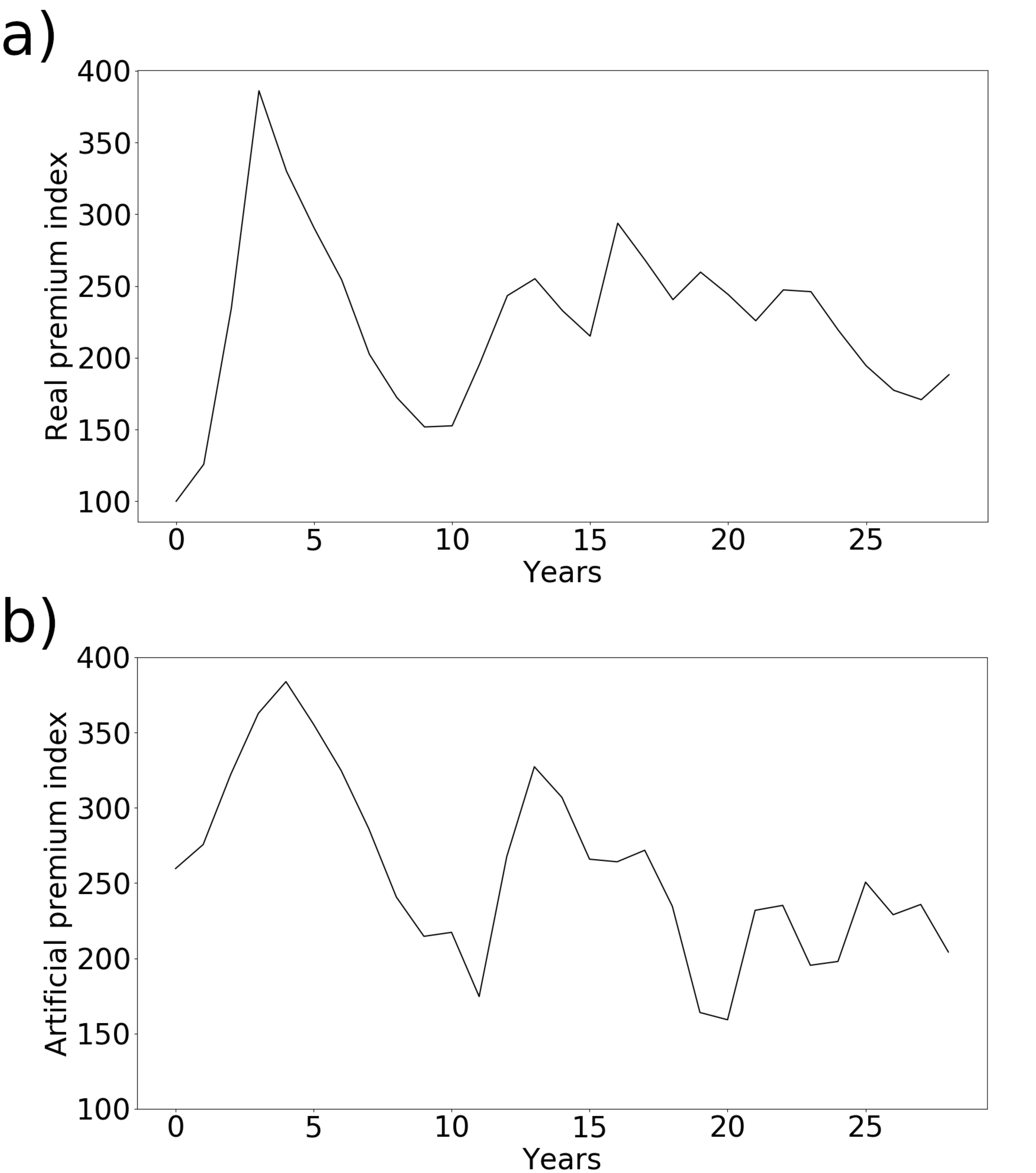}
  \caption{a) Real insurance cycle in terms of the \textit{Rate-on-Line index} (same as Figure~\ref{fig:global_rate_on_line}).
b) Insurance cycle as generated by the ABM, re-scaled to the same magnitude as in panel (a) and over a shorter span of tie in comparison to the previous figure.  We had to rescale because the algorithm used for computing this is not public, but the range of variation of the period and amplitude are similar.}
  \label{fig:comparison_index}
\end{figure}

The insurance model presented here is able to reproduce the most important stylized facts of the insurance cycle as discussed in section~\ref{sect:literature:insurancecycle}. In panel $a)$ of Figure~\ref{fig:combined}, the time evolution of the premium in a single run of the model over a span of more than $80$ years is shown. For the sake of simplicity we ran this simulation without reinsurance. The transitions from soft markets to hard markets can take several years. In line with real insurance cycles, fluctuations are irregular in both frequency and amplitude. The time series of profits and losses of the industry in the same simulation run is shown in panel $b)$ of Figure~\ref{fig:combined}. During most years, the industry is growing (profits are positive), but this growth is disrupted in years with catastrophes and the immediately following years.  The industry as a whole experiences losses only in years with catastrophes.

In Figure \ref{fig:comparison_index} we show a comparison between the real insurance cycle published by the reinsurance broker \textit{Guy Carpenter} and a simulated insurance cycle generated by the model.\footnote{Source:  \texttt{http://www.guycarp.com/}. 
We only have ca. 25 years of data since time series are only available starting in 1990.} The cycle generated by the model is a 25 year sample in a single run of more than 200 years. The algorithm Guy Carpenter uses to generate the index is not public as it is commercially sensitive information. To obtain a comparable measure, we have re-scaled the premium time series produced by the ABM to obtain units of the same magnitude as in the real index. Both time series share approximately the same range of variation in period and amplitude. 


\begin{figure}[tbhp]
  \centering
  \includegraphics[width=1\textwidth]{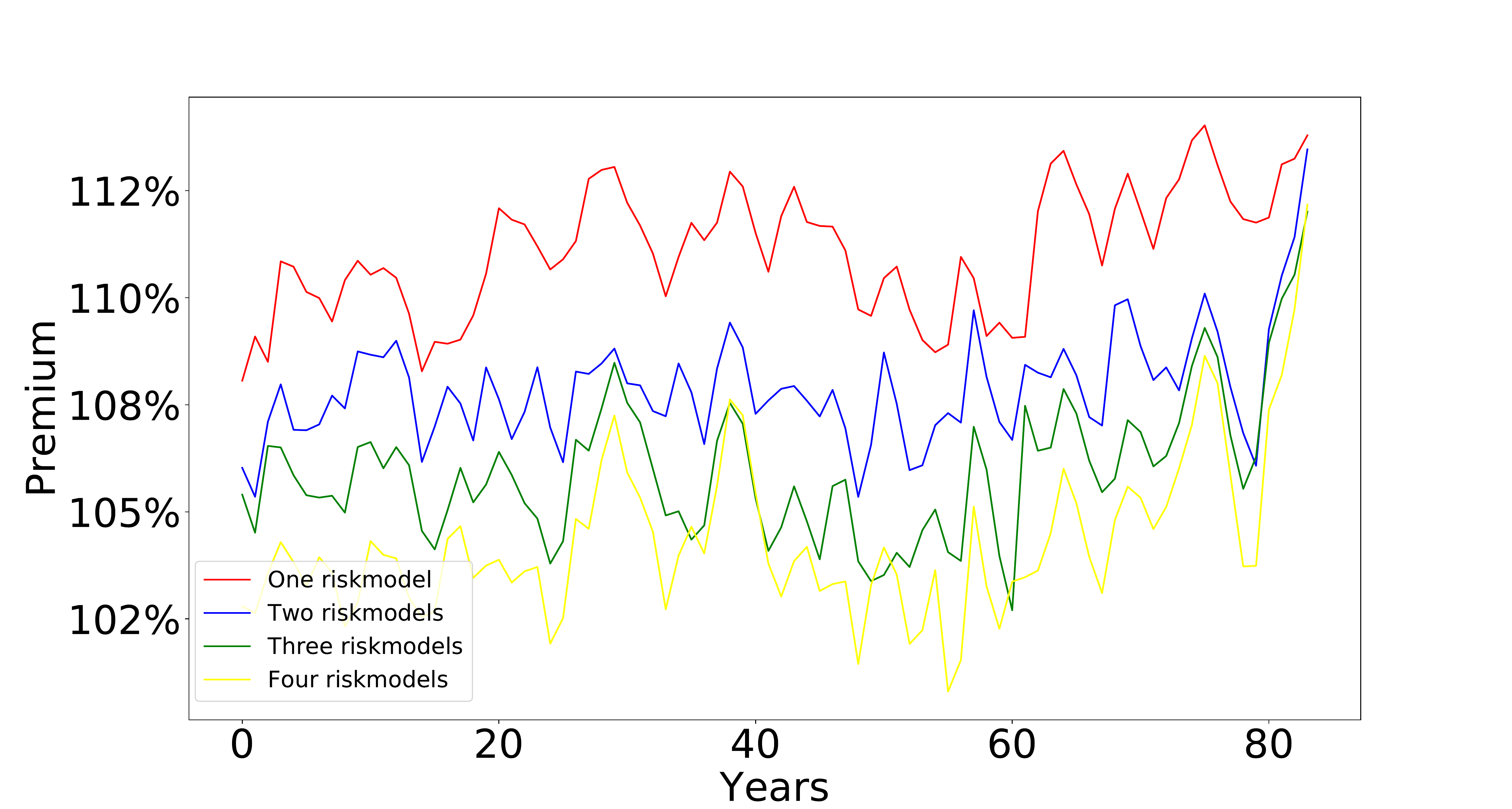}
  \caption{{\it Comparison of insurance cycles} resulting with identical risk events in different risk model diversity settings. The insurance cycle seems to be longer in the case of one risk model. The volatility/capital ratio is similar in all cases.}
  \label{fig:comparison_cycles}
\end{figure}

\begin{figure}[tbhp]
  \centering
  \includegraphics[width=.8\textwidth]{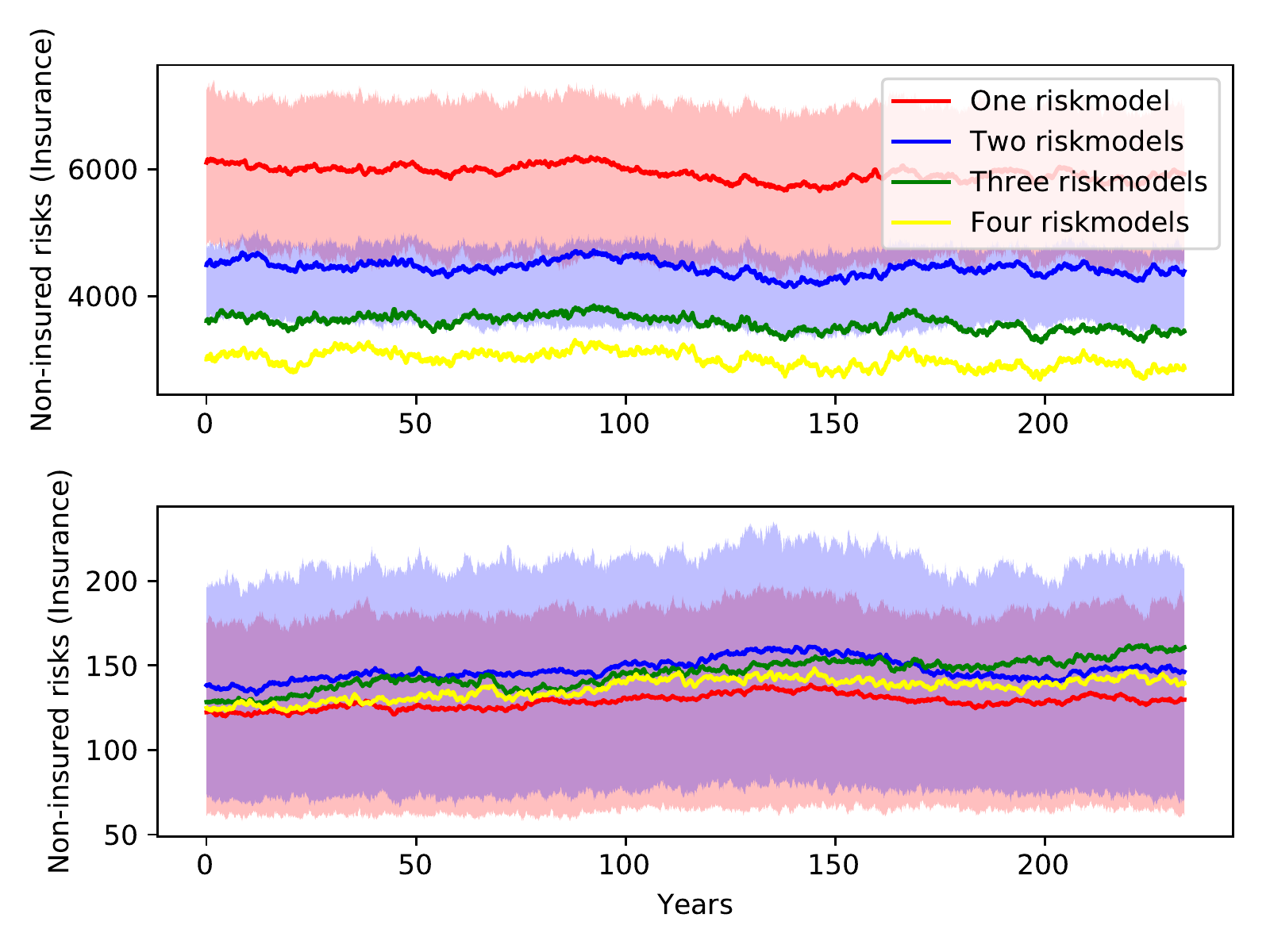}
  \caption{{\it Number of non-insured risks as a function of time.} Ensemble simulation with $400$ replications for each risk model diversity settings. Margin of safety is $\mu=2$. Time steps $1,200$-$4,000$ (months) are shown (transient in time setps $1$-$1,199$ removed). Ensemble means are shown as solid lines. The interquartile ranges of the settings with one (red) and two (blue) risk models are depicted as shaded areas. (The overlap of both areas is shaded in magenta.) 
  }
  \label{fig:time_mu2_re_contracts}
\end{figure}

\begin{figure}[tbhp]
  \centering
  \includegraphics[width=.8\textwidth]{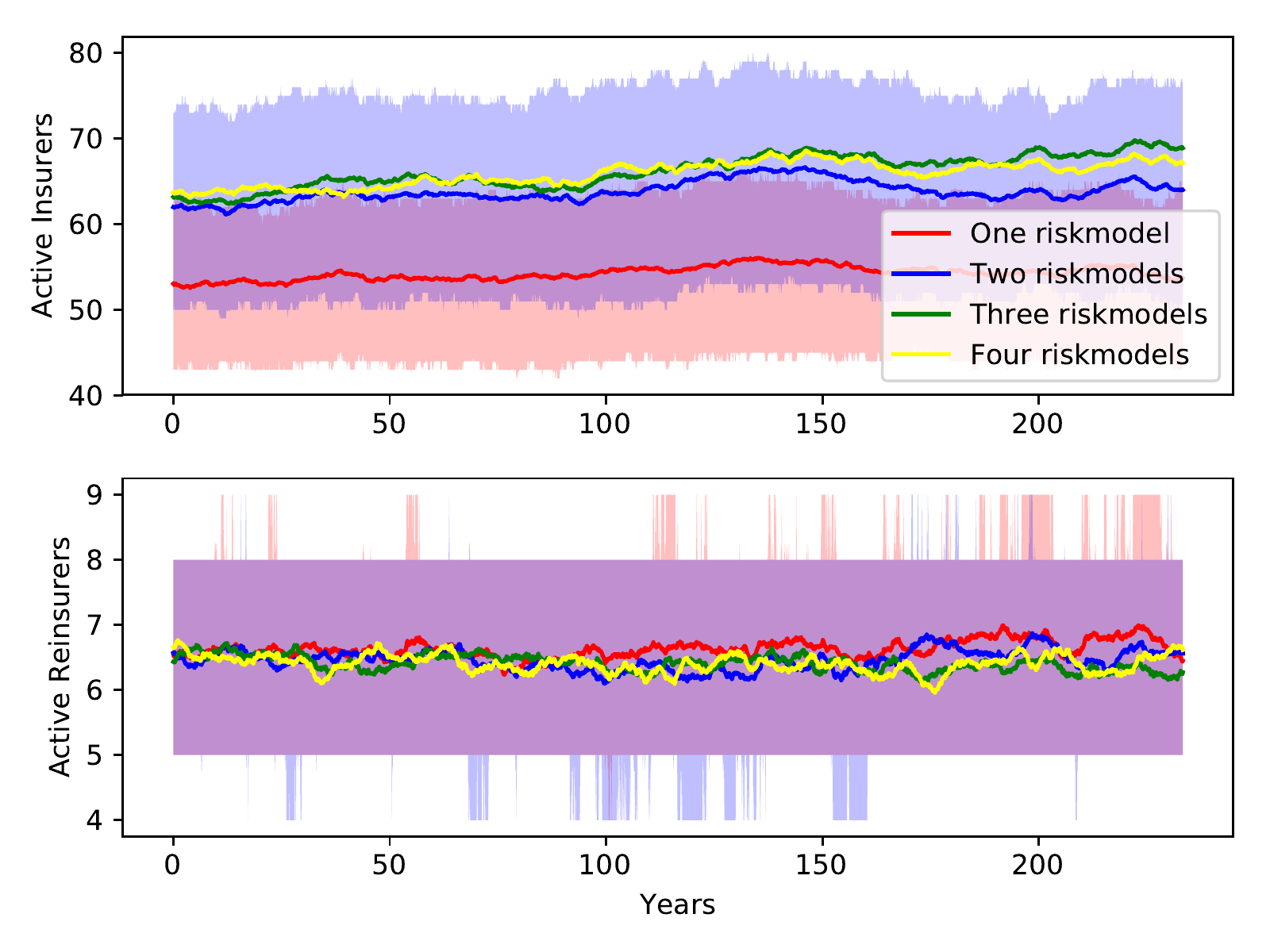}
  \caption{{\it Number of operational insurance firms.}  See caption for Figure~\ref{fig:time_mu2_re_contracts}.
}
  \label{fig:time_mu2_re_operational}
\end{figure}

\begin{figure}[tbhp]
  \centering
  \includegraphics[width=.8\textwidth]{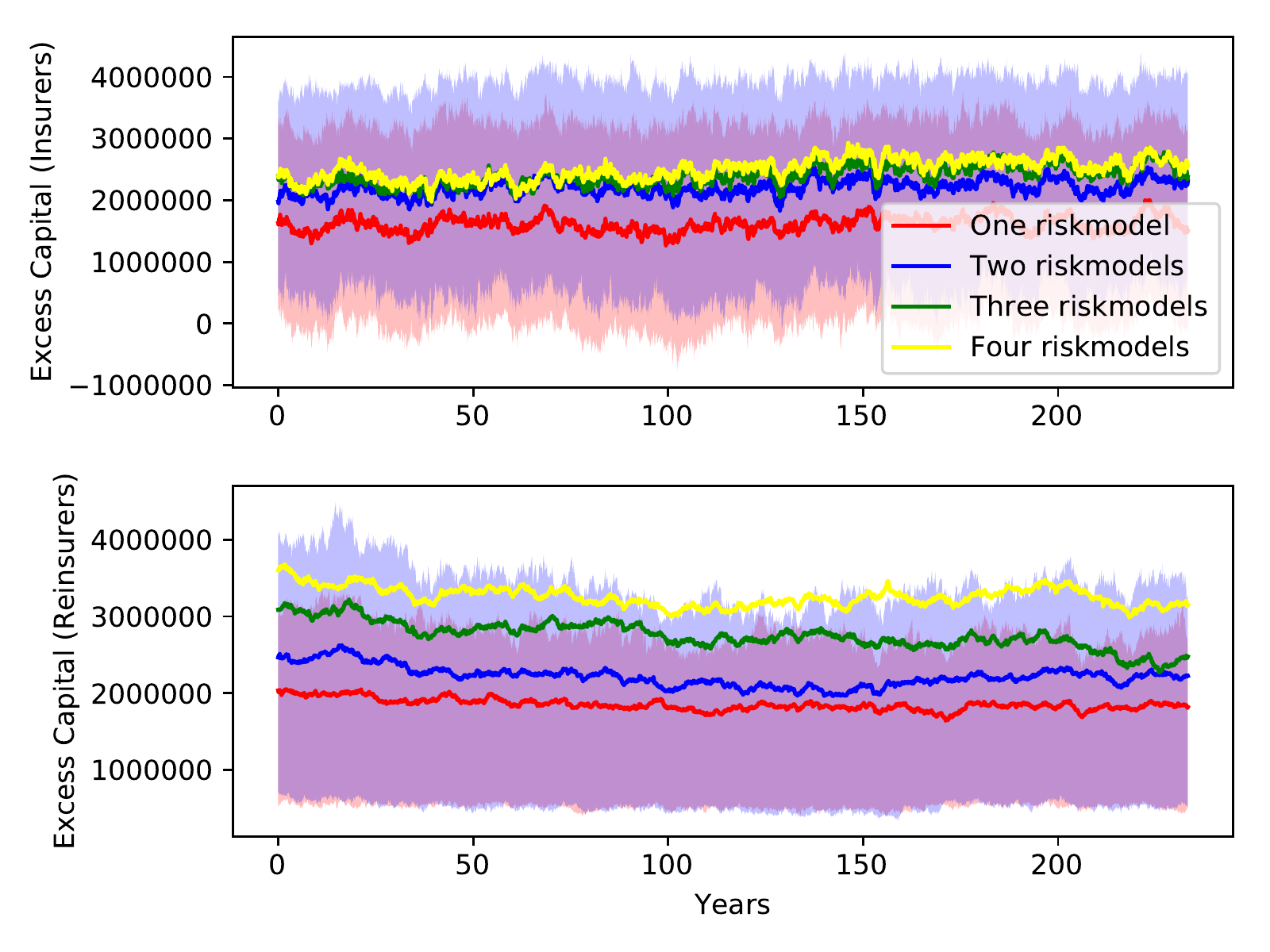}
  \caption{{\it Amount of excess capital} (beyond the capital required to cover currently underwritten contacts).  This provides a measure of of the capacity to write additional business. 
  See caption for Figure~\ref{fig:time_mu2_re_contracts}.
}
  \label{fig:time_mu2_re_excap}
\end{figure}

\begin{figure}[tbhp]
  \centering
  \includegraphics[width=.8\textwidth]{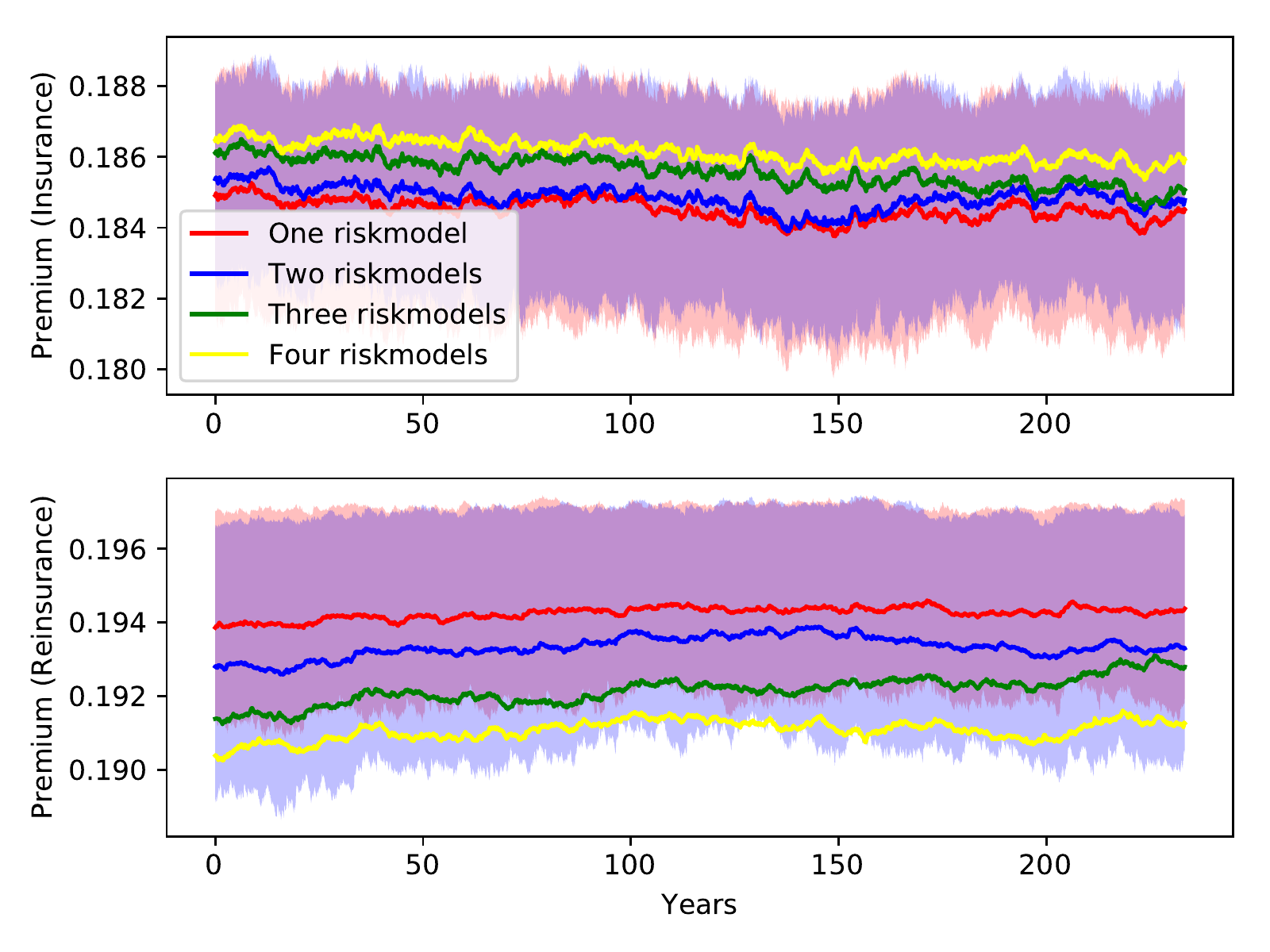}
  \caption{{\it Insurance premiums.}  See caption for Figure~\ref{fig:time_mu2_re_contracts}.
}
  \label{fig:time_mu2_re_premium}
\end{figure}

\subsection{Systemic risk due to model homogeneity} 
\label{sect:results:risk-model-diversity}


We now study the characteristics of the insurance system as we vary the number of distinct risk models available to the insurance and reinsurance companies from absolute homogeneity (one risk model) to four risk models with intermediate cases of two and three alternatives. Figure~\ref{fig:comparison_cycles} shows how the insurance cycle is affected by this in a simulation with the same schedule of catastrophes and random seed for all four risk model diversity settings. The premium tends to be lower when more risk models are available.  We have also find that the volatility/capital ratio is similar in all cases. The insurance cycle seems to be longer in the case of one risk model.  

As already mentioned, we have chosen the models so their average accuracy is the same but they make mistakes in different circumstances.  Specifically, each model underestimates risks in a different peril region. A catastrophe in a particular peril region will therefore hit firms that employ the one risk model which underestimates this peril region particularly hard. 
We perform 400 simulations for each of the four possible numbers of models and compute the means and the distributions of the behavior in each case. 

 The results for any given simulation are very diverse, with large variations from run to run.
 By performing 400 simulations we reduce the variation sufficiently to make the differences clear.  To reduce the variance we construct the $M=400$ ensembles for each of the four risk model diversity settings so that they have identical risk event profiles.\footnote{
That is, we consider $400$ different realizations of the stochastic processes governing when and where catastrophic perils of what size occur and run these realizations for each of the risk model settings. Initializing the simulation with the same emsemble of random seeds would not give similar risk event profiles and would not be a sufficiently meaningful comparison. As many aspects of the system are subject to heavy-tailed distributions, individual realizations might dominate the ensemble and bias the comparison if not present in the other three ensembles for the other risk model diversity settings.}  
The results corresponding to the four settings are shown in different colors in Figures~\ref{fig:time_mu2_re_contracts} through \ref{fig:time_mu2_re_premium}  in this section: red for the setting where all firms use the same risk model, blue for the setting with two different risk models, green for the setting with three risk models and yellow for the setting with four risk models.

%


As shown in Figure~\ref{fig:time_mu2_re_contracts}, the setting with one risk model typically results in more risks left without insurance coverage.  Since the number of insurable risks is held constant, if fewer contracts are issued, there are more risks that cannot be insured because no insurance firm is willing to insure them. This is mainly due to the fact that insurance firms find more difficult to diversify their portfolio with only one risk model hence they are more reluctant to issue more contracts.  When only one risk model is used, the number of risks without coverage is about $50\%$ higher on average than it is when four risk models are used.

The market is more competitive with a higher diversity of risk models: Figure \ref{fig:time_mu2_re_operational} shows that the number of insurance firms is increased from an average of 52 firms to around 68 when more models are used, an increase of $30\%$. Surprisingly, there is very little change in the number of reinsurers.  
As shown in Figure~\ref{fig:time_mu2_re_excap}, this also results in a reduction in the amount of available capital\footnote{Available capital is the capital that is not tied up covering existing contracts, essentially the capacity to write additional business.} for both insurance and reinsurance firms. 
Here the change is more dramatic:   For insurance companies, the available capital when there are four risk models is by a factor of about $1.5$ higher than it is with one risk model. For reinsurance companies, available capital it is higher by a factor of $1.75$.  This indicates that risk buffers are higher; companies are able to absorb more catastrophic loses with more model diversity.   

Finally, Figure~\ref{fig:time_mu2_re_premium} shows that for insurance firms the risk premiums are lower with one risk model than they are for four risk models, though here the difference is small (only $1$ percent).  Surprisingly, this effect is reversed for reinsurance firms, though once again, the difference is similarly small. The premium for reinsurance firms in the case of the setting with four risk models is around $2.5$ percent cheaper than the setting with only one risk model.

%

\begin{figure}[tbhp]  
  \centering
  \includegraphics[width=.8\textwidth]{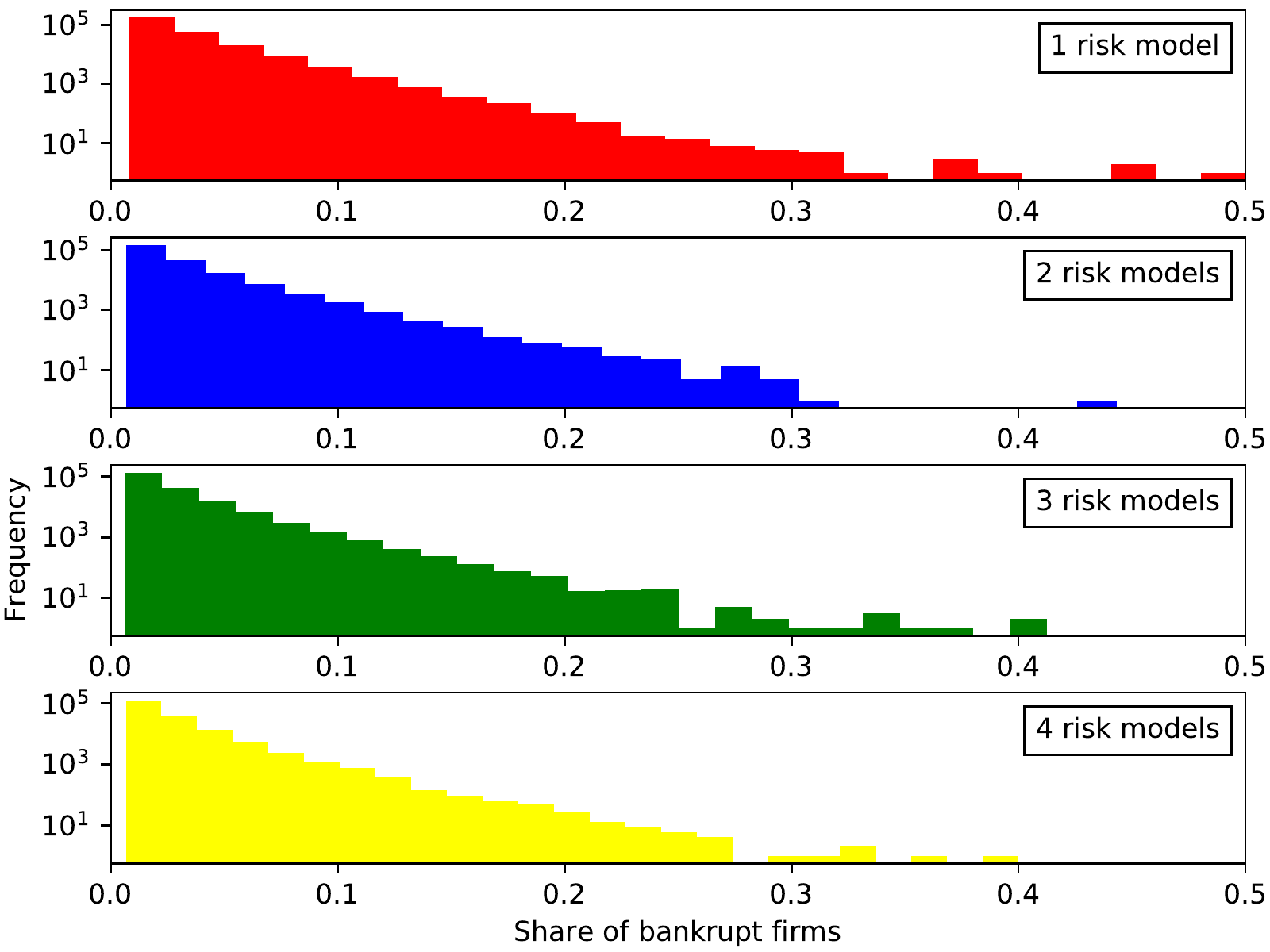}
  \caption{{\it Histogram of the total sizes of bankruptcy events}, measured as the fraction of firms $B$ that fail during each event.  ensemble of $400$ replications of simulations of $4,000$ time steps with margin of safety $\mu=2$. 
  The y-axis is in log scale.}
  \label{fig:hist_b_mu2_re}
\end{figure}
\begin{figure}[tbhp]
  \centering
  \includegraphics[width=.8\textwidth]{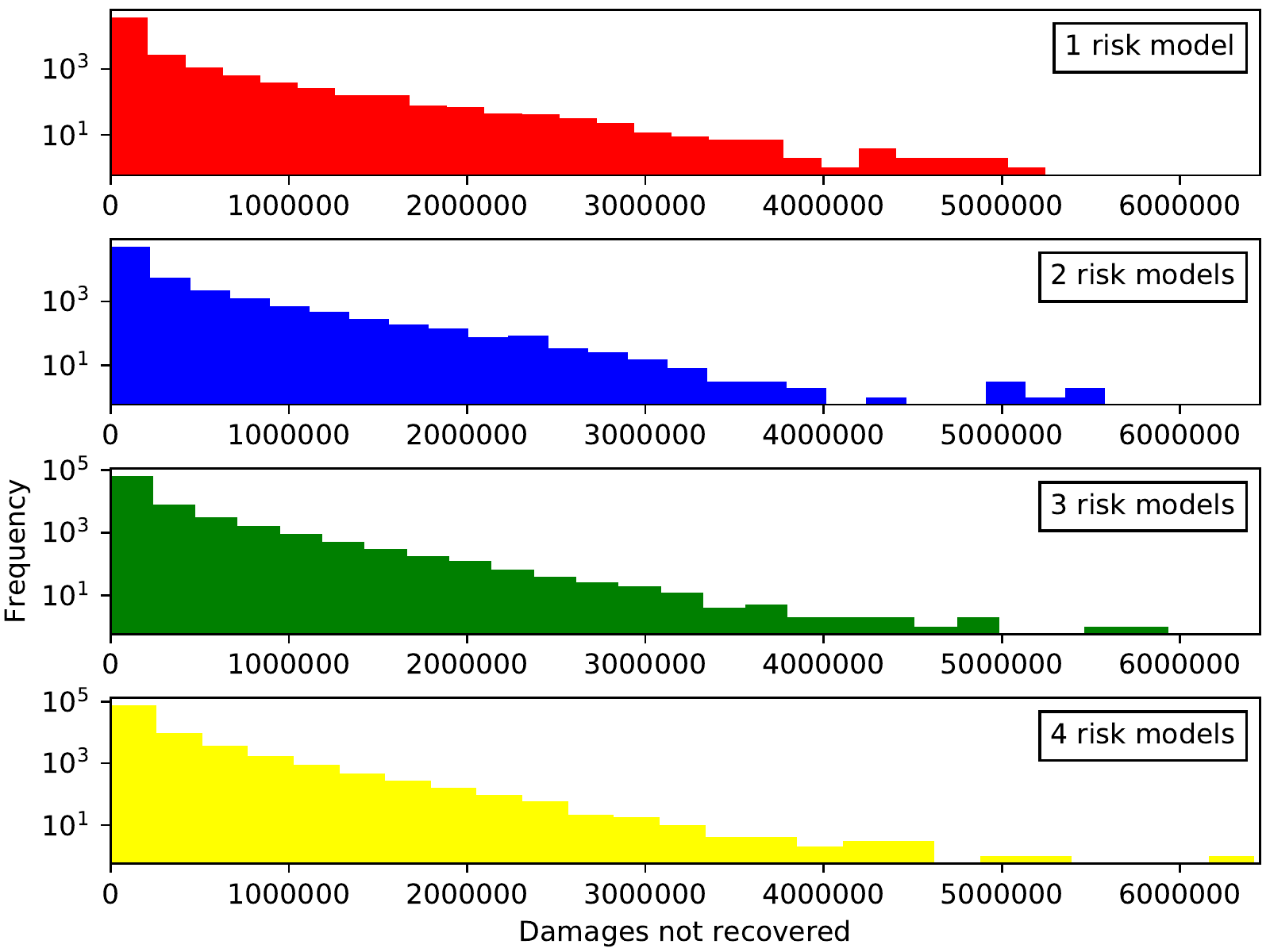}
  \caption{{\it Histogram of the number of non-recovered claims $C_t$ in each bankruptcy cascade}, where $C_t$ is defined on each timestep $t$.   See caption for Figure~\ref{fig:hist_b_mu2_re}. }
  \label{fig:hist_u_mu2_re}
\end{figure}


Bankruptcies are a key measure of systemic risk.  To study how the number of risk models affects this, we compile statistics about the {\it size of bankruptcy cascades}, which we measure as the share of bankrupt firms $B_{t}=b_{t}/f_{t}$. Here, $b_{t}$ is the number of bankrupt firms at time $t$ and and $f_{t}$ the total number of firms at time $t$).  Similarly, we look at the {\it number of non-recovered claims}.  This is the number of times that a policy isn't paid due to the default of an insurance company.  We measure this in terms of the number of unpaid claims $C_{t}$ at each time step.
Both numbers include both insurance and reinsurance firms. We study the distribution of these variables across all replications and the entire history of each replication of every setting of the simulation. Figure~\ref{fig:hist_b_mu2_re} shows the distributions of sizes of bankruptcy cascades\footnote{To account for indirect effects, a bankruptcy cascade is here defined as series of defaulting firms in successive time steps without intermediating time steps in which no bankruptcies occur.} while Figure~\ref{fig:hist_u_mu2_re} shows the distribution of the number of non-recovered claims. 

The total number of bankruptcy events seen in Figure~\ref{fig:hist_b_mu2_re} does not differ very much between the four risk model diversity settings.  However, the number of very large events is very different.  For the one risk model case (uppermost panel, red), the body of the distribution extends continuously up to more than a third of the sector ($0.35$) while in the four risk model case (lowermost panel) only some scattered outliers beyond $0.27$ are observed across all 160,000 time steps from all 400 replications.  As seen in Table~\ref{tab:lambdas}, roughly 4,200 firms default with one risk model, whereas there are only about $1,500$ firms defaulting with four risk models.  A similar, albeit less pronounced, picture emerges for the amounts of non-recovered claims (Figure~\ref{fig:hist_u_mu2_re}).

\begin{figure}[tbhp]
  \centering
  \includegraphics[width=.8\textwidth]{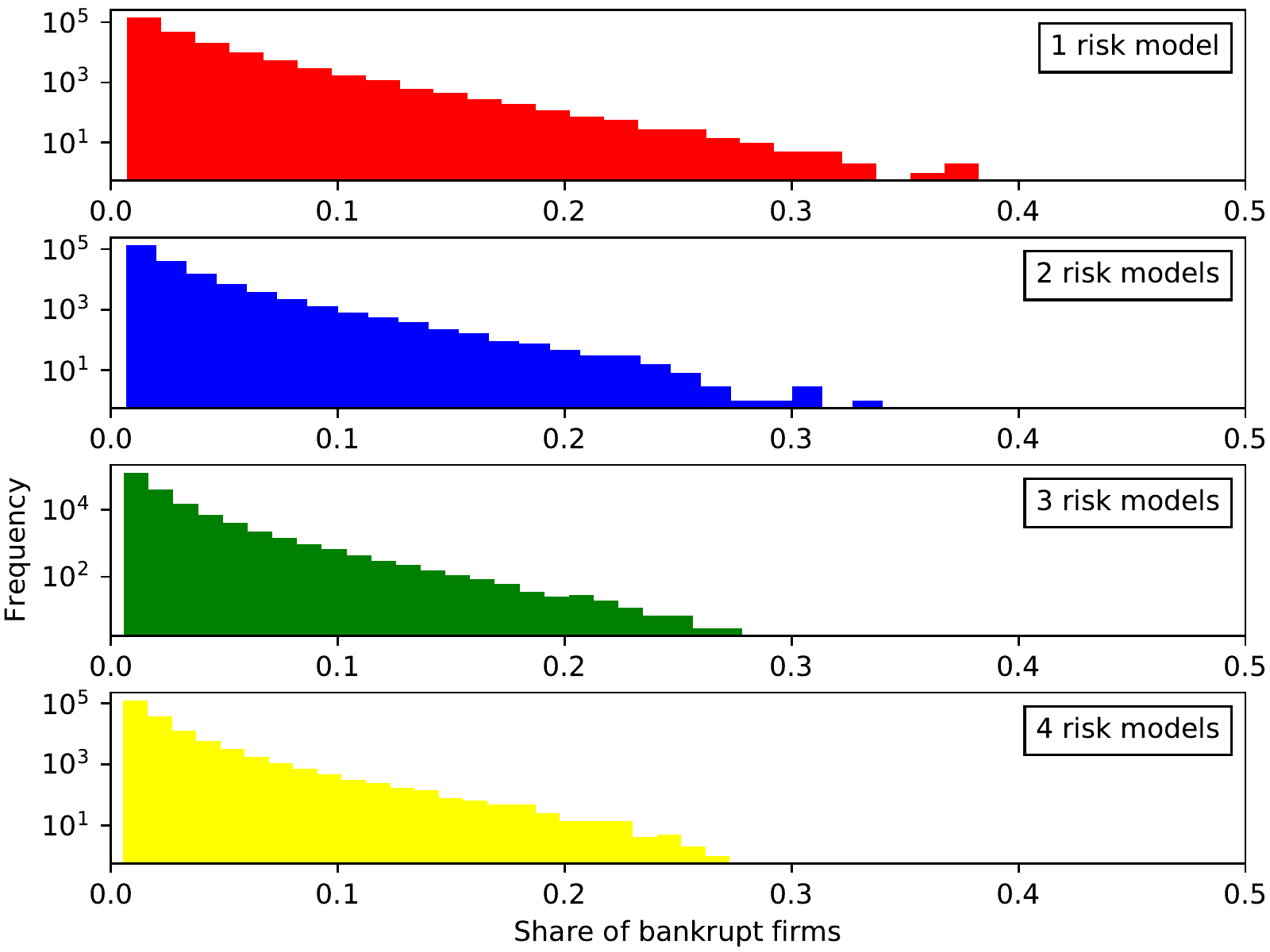}
  \caption{{\it Histogram of the total sizes of bankruptcy events without reinsurance.}   See caption for Figure~\ref{fig:hist_b_mu2_re}.  }
  \label{fig:hist_b_mu2_nore}
\end{figure}

The linearity of these histograms on semi-log scale suggests that exponentials provide a crude fit to the body of these distributions.  As an alternate measure of systemic risk we fit exponentials to each distribution \footnote{
The close-to-linear shape of the distributions in the semi-log plots in Figures~\ref{fig:hist_b_mu2_re} and \ref{fig:hist_u_mu2_re} suggests long-tailed distributions from the exponential family or similar. However, the exponential form must necessarily be truncated as the observed variable are shares that must be between $0$ and $1$.}
to measure the slope $\widehat{\lambda}$ with which the distribution decays in these semi-log plots.  Lower values indicate higher risk of very large events.   As shown in as shown in Table~\ref{tab:lambdas}, we find that the slope for the distribution of sizes of bankruptcy cascades $B$ is steeper for settings with more diversity, changing from $\widehat{\lambda} = 119$ with one risk model to $\widehat{\lambda} = 149$ with four risk models.\footnote{The effect is less clear-cut for the amount of unrecovered claims $C$. For instance, $C$ is lower for the setting with just one risk model, since the size of the insurance business is smaller, but the expected shortfall is larger.} This finding is robust and holds throughout four different series of simulations (each with all four risk model diversity settings) reported in the table: (1) the standard case, (2) a comparative case without reinsurance but all other settings identical (discussed in Section~\ref{sect:results:reinsurance}), (3) a comparative case with lower margin of safety ($\mu=1$), and (4) a case with lower margin of safety and without reinsurance (discussed in Subsection~\ref{app:sensitivity}). More diversity thus leads to many less events in specific tail quantiles than comparative settings with less diversity. This is confirmed by the number of bankruptcy events affecting more than 10\% of the insurance and reinsurance firms as reported in the lower part of Table~\ref{tab:lambdas}. The difference between the four risk model diversity settings becomes larger in cases without reinsurance. This is discussed in more detail in Subsection~\ref{sect:results:reinsurance} (compare Figure~\ref{fig:hist_b_mu2_nore}).

\begin{table}[tbhp]
\centering
\begin{tabular}{|p{4cm} | p{2cm} | p{2cm} | p{2cm} | p{2cm} |}
\hline\hline
    & \multicolumn{4}{c|}{\textbf{Parameter settings}}\\
\hline
\textbf{Margin of safety} & \textbf{$\mu=2$} &  \textbf{$\mu=2$} &  \textbf{$\mu=1$} &  \textbf{$\mu=1$} \\
\hline
\textbf{Reinsurance}      & \textbf{yes}     &  \textbf{no}      &  \textbf{yes}     &  \textbf{no}      \\
\hline
\textbf{Figure}           & \ref{fig:hist_b_mu2_re}     &  \ref{fig:hist_b_mu2_nore}      &  \ref{fig:hist_b_mu1_re}     &  \ref{fig:hist_b_mu1_nore}      \\
\hline\hline
    & \multicolumn{4}{c|}{\textbf{Slope $\widehat{\lambda}$ for \textit{sizes of bankruptcy cascades} ($B$)}}\\
\hline
One risk model    & 119 & 138 & 60 & 72 \\
Two risk models   & 145 & 151 & 65 & 83 \\
Three risk models & 154 & 173 & 65 & 87 \\
Four risk models  & 149 & 181 & 67 & 91 \\
\hline\hline
    & \multicolumn{4}{c|}{\textbf{Number of events with $>10\%$ of firms defaulting}}\\
\hline
One risk model    & 4212 & 4385 & 22486 & 21928 \\
Two risk models   & 3013 & 2453 & 16137 & 12699 \\
Three risk models & 1981 & 1686 & 12419 & 7952 \\
Four risk models  & 1561 & 1229 & 10323 & 5441 \\
\hline\hline
\end{tabular}
\caption{Downward slopes $\widehat{\lambda}$ of the distributions of the {\it sizes of bankruptcy cascades}  ($B$)
, obtained from exponential fit and numbers of events in the right tail beyond $10\%$ of all firms bankrupt.}
\label{tab:lambdas}
\end{table}


\begin{figure}[tbhp]
  \centering
  \includegraphics[width=.8\textwidth]{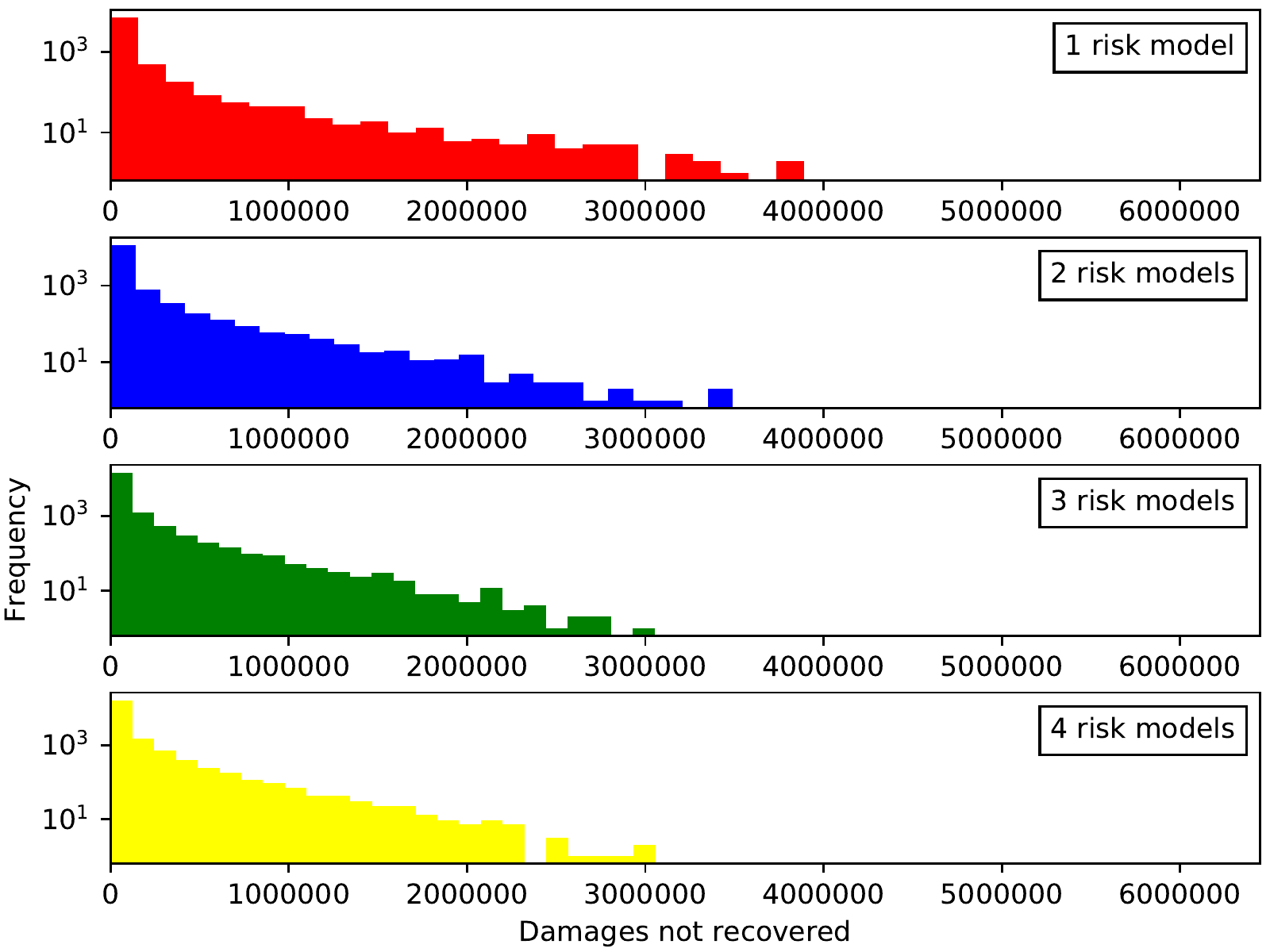}
  \caption{{\it Histogram of the number of non-recovered claims $C_t$ at each time step when reinsurance is used.}     See caption for Figure~\ref{fig:hist_b_mu2_re}. }
  \label{fig:hist_u_mu2_nore}
\end{figure}

\begin{figure}[tbhp]
  \centering
  \includegraphics[width=.8\textwidth]{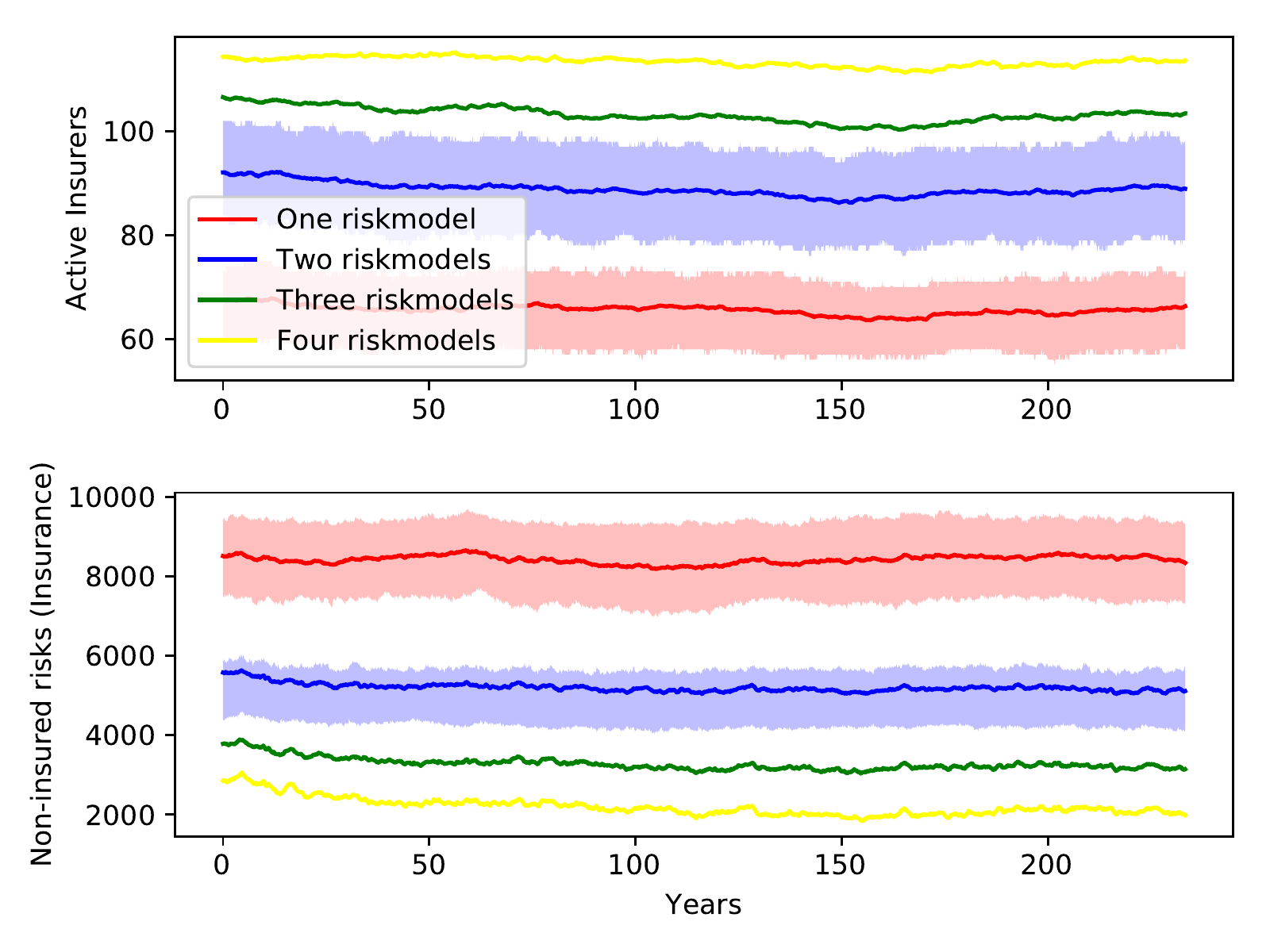}
  \caption{{\it Number of operational firms and number of insured risk without reinsurance.}  See caption for Figure~\ref{fig:time_mu2_re_contracts}.
}
  \label{fig:time_mu2_nore_contracts}
\end{figure}

\begin{figure}[tbhp]
  \centering
  \includegraphics[width=.8\textwidth]{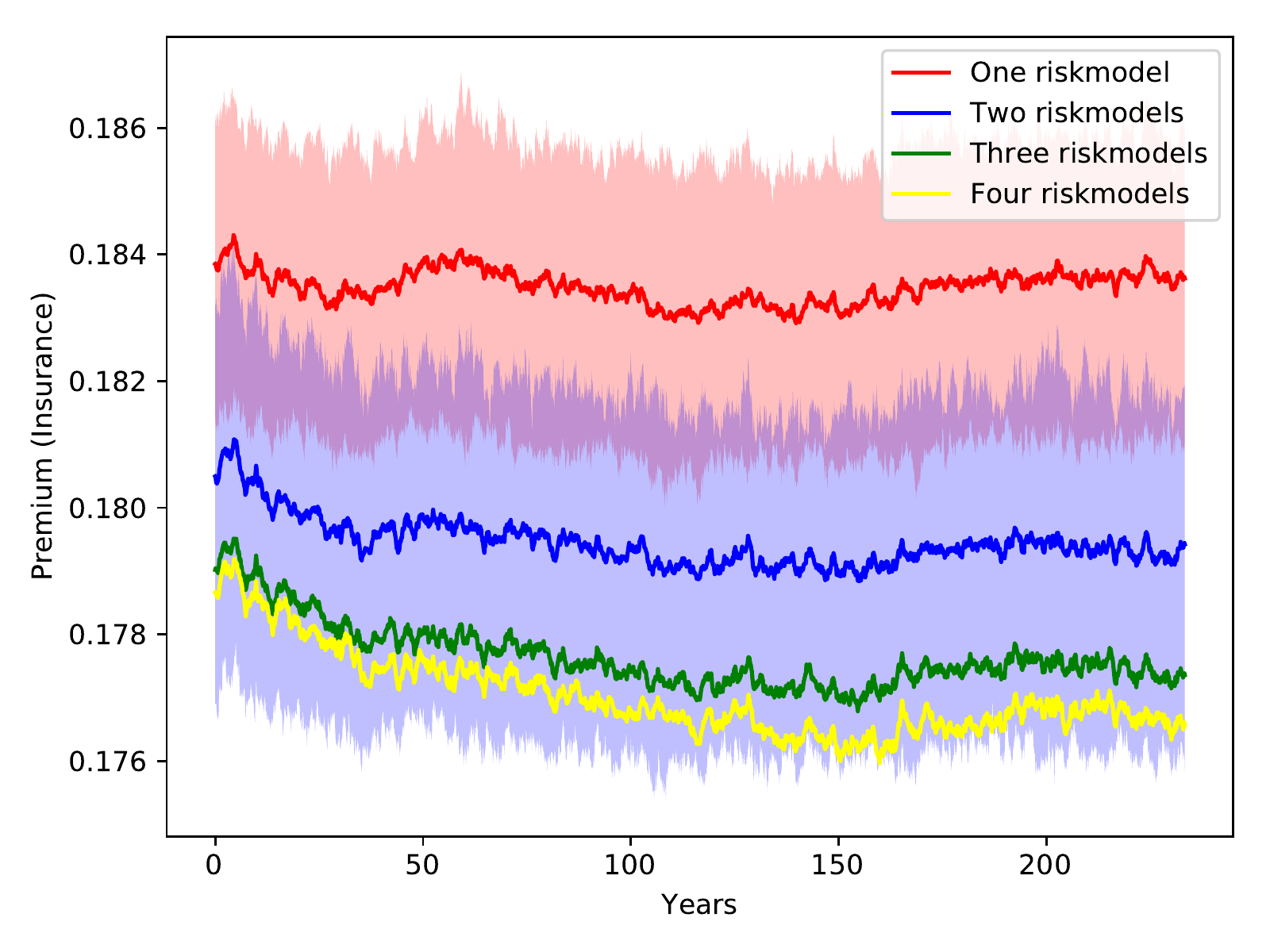}
  \caption{ {\it Insurance premiums without reinsurance.}  See caption for Figure~\ref{fig:time_mu2_re_contracts}.
}
  \label{fig:time_mu2_nore_premium}
\end{figure}


\subsection{Effect of reinsurance}
\label{sect:results:reinsurance}

The effect of reinsurance can be observed by running the simulation without reinsurance firms. The results are reported in Figures \ref{fig:hist_b_mu2_nore} (histogram of sizes of bankruptcy cascades), \ref{fig:hist_u_mu2_nore} (histogram of amounts of non-recovered claims), as well as \ref{fig:time_mu2_nore_contracts} and \ref{fig:time_mu2_nore_premium} (time series). 
It can be seen that the effect of risk model diversity or homogeneity on bankruptcy cascades is much stronger in this case.  For example, 
the shapes of the distributions for the four risk model settings are markedly different in the case without reinsurance (Figure~\ref{fig:hist_b_mu2_nore}) with the tail becoming shorter for settings with more diversity. With only one risk model, there were 4385 events in the sample that affected more than 10\% of the insurance firms. With four risk models, this reduces to 1229 events, a reduction of 71\% (see Table~\ref{tab:lambdas}, second column). In the equivalent settings with reinsurance (Figure~\ref{fig:hist_b_mu2_re}), this reduction is only 62\% from 4212 to 1561 events. The difference is also clear from the slope of the fit line of the histograms of bankruptcy sizes in semi-log (that is, the parameter of the exponential distribution, $\widehat{\lambda}$): Without reinsurance it changes from $\widehat{\lambda}=181$ to $\widehat{\lambda}=149$, a change of 21\%, with reinsurance only 16\% from $\widehat{\lambda}=149$ to $\widehat{\lambda}=119$ (see Table~\ref{tab:lambdas}, first and second column).

While reinsurance adds a second contagion channel to systemic risk due to the counterparty exposure from reinsurance contracts, it partially alleviates the systemic effects of risk model homogeneity. Therefore, the number of large bankruptcy events (more than 10\% of firms affected) is up to 20\% higher with reinsurance. This can be seen in the first and second column in Table~\ref{tab:lambdas}. It holds for all settings with at least 2 risk models. In the setting with risk model homogeneity (only one risk model) this is reversed and the number of large bankruptcy events is larger without than with reinsurance. The increased effect of risk model homogeneity with reinsurance discussed in the previous paragraph, is in this setting stronger than the counterparty exposure effect.

The stronger effects of risk model homogeneity are also visible in the time series shown in Figures~\ref{fig:time_mu2_nore_contracts} and \ref{fig:time_mu2_nore_premium} (compared to Figures~\ref{fig:time_mu2_re_operational}, \ref{fig:time_mu2_re_contracts}, and \ref{fig:time_mu2_re_premium}): The ensemble means for the different risk model diversity settings lie further apart: With reinsurance, risk model diversity can increase the number of active insurers from about 50 to about 65 (Figure~\ref{fig:time_mu2_re_operational}) and reduce the number of uninsured risks from about 6000 to half that (Figure~\ref{fig:time_mu2_re_contracts}). Without reinsurance, risk model diversity can increase the number of active insurers from about 70 to about 120 and reduce the number of uninsured risks from about 8000 to about 2000 (Figure~\ref{fig:time_mu2_nore_contracts}). The effect on the ensemble mean of the premium is also about 50\% stronger without reinsurance. Moreover, while the ensemble interquartile ranges overlap for each of these variables with reinsurance, this overlap is not present without reinsurance, except in the premium prices, and there the overlap is small.
%

It should be noted, however, that the numbers of uninsured risks are higher without than with reinsurance for almost all risk model diversity settings. This emphasizes that reinsurance does have a productive and important role in the insurance system beyond rearranging the patterns of systemic risk of modeling.

\begin{figure}[tbhp]
  \centering
  \includegraphics[width=.9\textwidth]{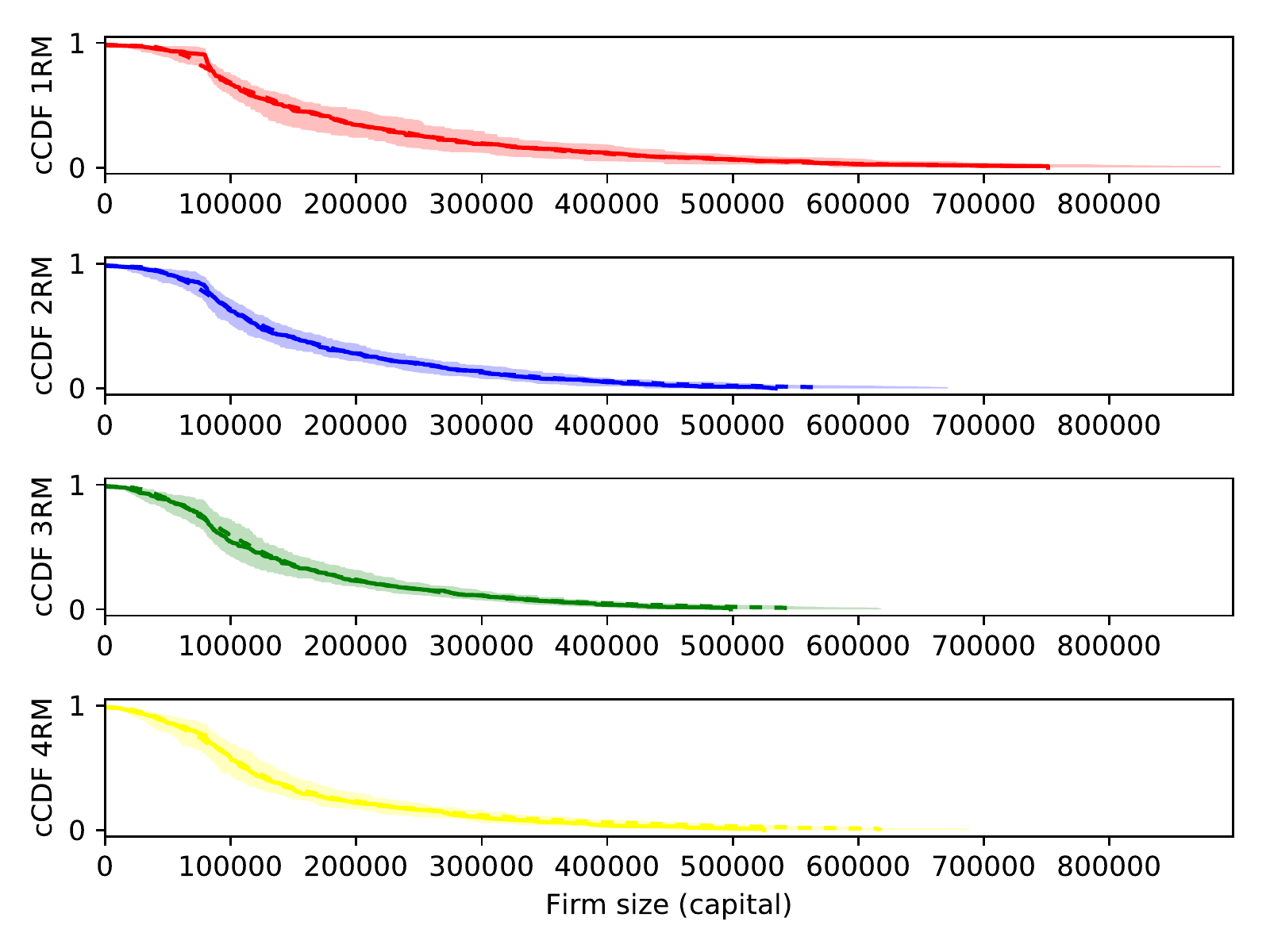}
  \caption{Ensemble of empirical complementary cumulative distribution functions (cCDFs) of the distribution of insurance firm sizes in terms of capital after $1,000$ time steps in an ensemble of $70$ replications of simulations with margin of safety $\mu=2$. The median is shown as solid line, the mean as dashed line, the interquartile range as shaded area. Mean, median, and interquartile range are with respect to the ensemble of cCDFs, i.e., evaluated in x-axis direction.}
  \label{fig:ccdf_i_mu2_re}
\end{figure}

%
%

\subsection{Reproducing the emergence of asymmetric firm-size distributions} 
\label{sect:results:distributions}


Figure~\ref{fig:ccdf_i_mu2_re} shows the complementary cumulative distribution function of the distribution of insurance firm sizes in terms of capital after 1000 time steps with dispersion in an ensemble run for one parameter setting. 
The number of firms ranges up to several hundred per replication. Distributions with a long tail emerge consistently across all risk model diversity settings. This is insensitive to parameters (e.g., $\mu$, presence or absence or reinsurance, etc.), and also appears for reinsurance firms (see Appendix \ref{app:sensitivity}). This corresponds nicely with established empirical facts about firm size distributions, which are found to be long tailed although the findings on the concrete functional form diverge;\footnote{The empirical distribution is even more heavy-tailed than that typically emerging in the model.} Lognormal, exponential, or power law shapes have been proposed. 
This fact can be confirmed for the insurance sector, but the number of firms, both in the simulation replications in our study and in empirical data sets on insurance firms, is not large enough to fit concrete functional forms with sufficient confidence.

\section{Conclusion}
\label{sect:conclusion}


Solvency II, the EU insurance regulation framework, came into effect in January 2016. It constitutes a major step for insurance regulation in terms of liquidity, capital, and transparency requirements, making it possible to address microprudential aspects as well as potentially systemic risk from counterparty exposure. It also includes provisions for usage and design of risk models. But whether we understand the systemic level of the insurance and reinsurance business sufficiently well to confidently design regulatory measures for a resilient insurance sectors is still an open question. No scholarly consensus has as yet emerged about what drives the insurance cycle. It remains unclear what effect new financial market vehicles like CAT bonds will have on the industry compared to traditional reinsurance. The investigation of systemic risk in insurance is a new and unexplored field. 

In the present paper we present an agent-based model of the insurance sector to help to address these questions. The model includes reinsurance and a number of other aspects. It reproduces a variety of stylized facts, ranging from the insurance cycle to the firm size distribution to the importance of reinsurance. It also allows investigating the roles of various elements of the insurance system and the mechanisms behind some of its characteristics. 

We have demonstrated the capabilities of the model to reproduce said stylized facts and used it to show the existence and the properties of systemic risk of modeling in insurance systems. To do so, we considered ensemble runs with the same environment, the same parameters, and the same profile of risk events but different numbers of (one, two, three, and four) risk models of identical quality employed by the insurance and reinsurance firms in the simulations. We found that settings with greater diversity tend to experience less severe bankruptcy cascades, especially in cases with a low margin of safety and in counterfactual cases without reinsurance. 

We found that settings with risk model diversity not only succeeded in partially offsetting the risk of large bankruptcy cascades, but also tended to lead to an insurance-reinsurance sector with greater penetration (higher share of risks underwritten), more active firms, and more available capital for additional endeavours on the part of the insurance firms. However, we found that the benefits differed between the insurance and the reinsurance part of the business; doubtlessly, different parameters can lead to a reallocation of assets between these sectors. Reinsurance tends to mitigate the strength of the systemic risk effect of risk model homogeneity but can exacerbate it in some cases by adding an additional contagion channel (reinsurance counterparty exposure).

It should be noted that the results reported here represent an entirely hypothetical world that was only calibrated in terms of accurate functioning of the interaction mechanisms and credible settings of the environmental parameters, such as the distributions of perils, the interest rate, the rate of market entry. Running simulations that are calibrated to empirical data of real insurance-reinsurance markets is highly desirable, but will require high quality data as well as significant efforts in model calibration,\footnote{Agent-based model calibration remains a very active field of research; sufficiently flexible and computationally feasible methods are recent and still under development \citep{Lampertietal17,Fagioloetal17,Barde/vanderHoog17,Platt19}.} and is left for future research.

\section*{Acknowledgements}
\label{sect:acknowledgements}

This work has been fully supported by MS Amlin under the project Systemic Risk of Modelling. We acknowledge enlightening discussions with JB Crozet, Trevor Maynard, and David Singh. We also acknowledge Sandtable for allowing us to use their platform Sandman to compute the ensemble of simulations required for this paper.

\bibliographystyle{apalike}
\bibliography{main}

\newpage
\appendix

\section{Standard parameter setting}
\label{app:parameters}

\begin{table}[h!]
  \centering
  \begin{tabular}{r l l }
\hline\hline
{\bf Symbol}& {\bf Variable} & {\bf Value} \\\hline\hline
$t_{max}$& Number of time steps & $4,000$ \\
$\mu$& Margin of safety & $2.0$ \\
$\alpha $ & VaR exceedance probability & $0.005$ \\
$\varrho$& Dividends as share of profit & $0.4$\\
$\xi$& Monthly interest rate & $0.001$\\
$M$& Number of replications per setting & $400$ \\
$\overline{k_i}$& Initial capital (insurance firms) & $80,000$\\
$\overline{k_r}$& Initial capital (reinsurance firms) & $2,000,000$\\
$f_{i,0}$& Initial number of insurance firms & $20$ \\
$f_{r,0}$& Initial number of reinsurance firms  & $4$\\
$\eta_{i,0}$& Insurance firm market entry rate & $0.3$ \\
$\eta_{r,0}$& Reinsurance firm market entry rate & $0.05$ \\
$\gamma_i$ & Capital employment threshold for insurance firm exit & $ 0.6 $ \\
$\tau_i$ & Time limit for insurance firm exit  & $ 24 $ \\
$\gamma_r$ & Capital employment threshold for reinsurance firm exit & $ 0.4 $ \\
$\tau_r$ & Time limit for reinsurance firm exit  & $ 48 $ \\
$\lambda$& Average frequency of perils (per peril region) & $0.03$ \\
$\sigma$& Tail exponent of damage distribution & $-2$ \\
$n$ & Number of peril regions & $4$ \\
$H$ & Number of risks & $20,000$ \\
$\zeta$ & Risk model inaccuracy & $2$ \\
$MinL$ & Lower premium limit factor & $70\%$ \\
$MaxL$ & Upper premium limit factor & $135\%$ \\
$s_i$ & Insurance premium sensitivity parameter & $1.29 \times 10^{-9}$ \\
$s_r$ & Reinsurance premium sensitivity parameter & $1.55 \times 10^{-9}$ \\

\hline\hline
\end{tabular}
  \caption{Standard parameter setting of the simulation}
  \label{tab:parameters}
\end{table}

\section{Technical description of the model}
\subsection{Catastrophes time distribution}\label{app:time_dist}

We assume that the number of catastrophes in the different peril regions follow a Poisson distribution, which means that the separation time between them is exponentially distributed with density function:
\begin{equation}
e(t) = \lambda e^{-\lambda t}.
\end{equation}
where $\lambda$ is the parameter of the exponential distribution and the inverse of the average time between catastrophes. We generally set $\lambda=3/100$, that is, a catastrophe should occur on average every $33$ years. We draw all the random variables necessary to set the risk event profile (when a catastrophe occurs and of what size the damages are) at the beginning of the simulation. In order to compare the $n$ different ``worlds'' with different risk model diversity settings, we set the same $M$ risk event profiles for the $M$ replications of all $n$ risk model diversity settings. That is, the same hypothetical ``worlds'' with different risk model settings, but with the same catastrophes happening at the same time and with the same magnitude, are compared. 

\subsection{Global loss distribution}\label{app:global_loss}

We use a Pareto probability distribution $\varphi$ for the total damage inflicted by every catastrophe, since historically they follow a power law. The Pareto distribution is defined as
\begin{equation}
\varphi(D_x) = \frac{\sigma}{D_x^{\sigma+1}},
\end{equation}
where $D_x$ are the values of the damages caused by the catastrophes. We generally set the exponent $\sigma=2$. The distribution is truncated with a minimum (below which the damage would be too small to be considered a catastrophic event) and a maximum. The maximum is given by the value of insured damages. The density function is therefore:
\begin{equation}
    \tilde{\varphi}({D_x}) = \begin{cases} 
      0 & 1 \leq D_x, \\
      \frac{\varphi(D_x)}{\int_{0.25}^{1} \varphi(D_x)d D_x}    & 0.25\leq D_x\leq 1, \\
      0 & D_x \leq 0.25.
   \end{cases}
\end{equation}
Like the separation times, the damages of the catastrophes are drawn at the beginning of the simulation and are the same for the different risk model settings. We denote the total normalized loses drawn form this truncated Pareto distribution as $L_i$. 

\subsection{Individual loss distribution}\label{app:individual_loss}

For the sake of simplicity we assume all risks in the region to be affected by the catastrophe, albeit with a different intensity. To determine the specific distribution of the known total damage across individual risks we use a beta distribution, defined as
\begin{equation}
\beta(d_x) = \frac{\Gamma(g+h)x^{g-1}(1-d_x)^{h-1}}{\Gamma(g)\Gamma(h)},    
\end{equation}
where $\Gamma$ is the Gamma function and $d_x$ is in this case the individual loss inflicted by the catastrophe to every individual risk. The two parameters $g$ and $h$ determine the shape of the distribution and define the expected value of the beta distribution which is,
\begin{equation}
E[\beta(x)] = \frac{g}{g+h}.
\end{equation}
Since the total loss inflicted by the catastrophe is $L_x$ and this should match the expected value (for large numbers of risks), we use this fact to compute $h$ for every catastrophe while always setting $g=1$. That is,
\begin{equation}
L_x = \frac{1}{1+h}.
\end{equation}
Solving for $b$ we get
\begin{equation}
h = \frac{1}{L_x} -1.
\end{equation}
The shape of the individual loss distribution depends on the total loss value and has to be adjusted for every catastrophe. We draw as many values from the distributions as we have risks in the peril region. Finally, the claims received by the insurer $j$ from all risks $i$ insured by $j$ are computed as
\begin{equation}
Claims_{x,j} = \sum_i \begin{cases} 
      min(e_i, d_{x,i} \cdot v) - Q_i & Q_i \leq d_{x,i} \cdot v_i, \\
      0 & d_{x,i} \cdot v_i \leq Q_i.   
   \end{cases}
\end{equation}
where $e_i$ is the excess of the insurance contract, $d_{x,i}$ is the individual loss, $v_i$ the total value of the risk and $Q_i$ is the deductible. For convenience, we generally have $v_i=1$.

\subsection{Dividends}\label{app:dividends}

Firms in the simulation pay a fixed share of their profits as dividends in every iteration, provided there were positive profits. In case time periods in which the firm writes loses no dividend is paid. That is,
\begin{equation}
        R = max(0, \varrho \cdot \mathrm{profits}), 
\end{equation}
where $R$ are the dividends and $\varrho$ is the share of the profits that is paid as dividends. For the results that we report in this paper we have fixed $\varrho=0.4$ for all the simulations done. 

\subsection{Pricing}\label{app:premiums}

Premiums in the model are global, but different for insurance and reinsurance contracts.

The insurance premium $p_t$ at time step $t$ depends on the total capital available in insurance at that time, $K^{T}_t$:
\begin{equation}\label{pricingeq}
    p_t = \begin{cases} 
      MaxL & MaxL\leq p_t \\
      p_f * MaxL - \frac{s \times K^{T}_t}{K^{I}_0 \times \widetilde{D} \times H}    & MinL\leq p_t\leq MaxL \\
      MinL & p_t\leq MinL,
   \end{cases}      
\end{equation}
where $MaxL$ is the maximum loading that the policyholders are willing to accept (upper bound) and $MinL$ is the min loading that the insurers are willing to consider in order to underwrite a policy (lower bound). The lower bound $MinL$ is smaller than $1$ since insurers usually underwrite premiums in soft market conditions in order to keep market share. Between the upper and lower bounds, the price is a falling linear function of the amount of capital with the intercept $p_f \times MaxL$ and slope $-\frac{s_i}{\overline{k}_i \times \widetilde{D} \times H}$ is applied. Here, $p_f$ is a standard premium that matches the expected losses plus a markup. The equation further depends on number of risks available in the market, $H$
, the expected damage by risk, $\widetilde{D}$, and the initial capital held by insurers at the beginning of the simulation, $\overline{k}_i$. Finally, $s_i$ is a sensitivity parameter.   

The reinsurance premium price equation differs from Eq.~\ref{pricingeq} in that only reinsurance capital is considered in $K^{T}_t$ and the sensitivity parameter $s_r$ is different. 

\subsubsection*{Stock-flow consistency}

As the model does not have a macro-economic perspective, stock-flow consistency requirements in the standard sense do not apply. However, the simulation model does not allow anything to appear from nothing or disappear into nothing. The rest of the economy is represented as a separate quasi-agent that handles all payments into or out of the insurance and reinsurance sector and is endowed with a very large but finite amount of capital. This is initialized at the beginning of the simulation and updated at every simulation step. In practice, dividend payments and claim payments to insurance customers are made to this agent while premiums from insurance customers are paid by this agent. This allows us, among other things, to keep track of the payment balance between the insurance sector and the rest of the economy.

\section{System sensitivity} 
\label{sect:results:sensitivity}
\label{app:sensitivity}

\begin{figure}[tbhp]
  \centering
  \includegraphics[width=.8\textwidth]{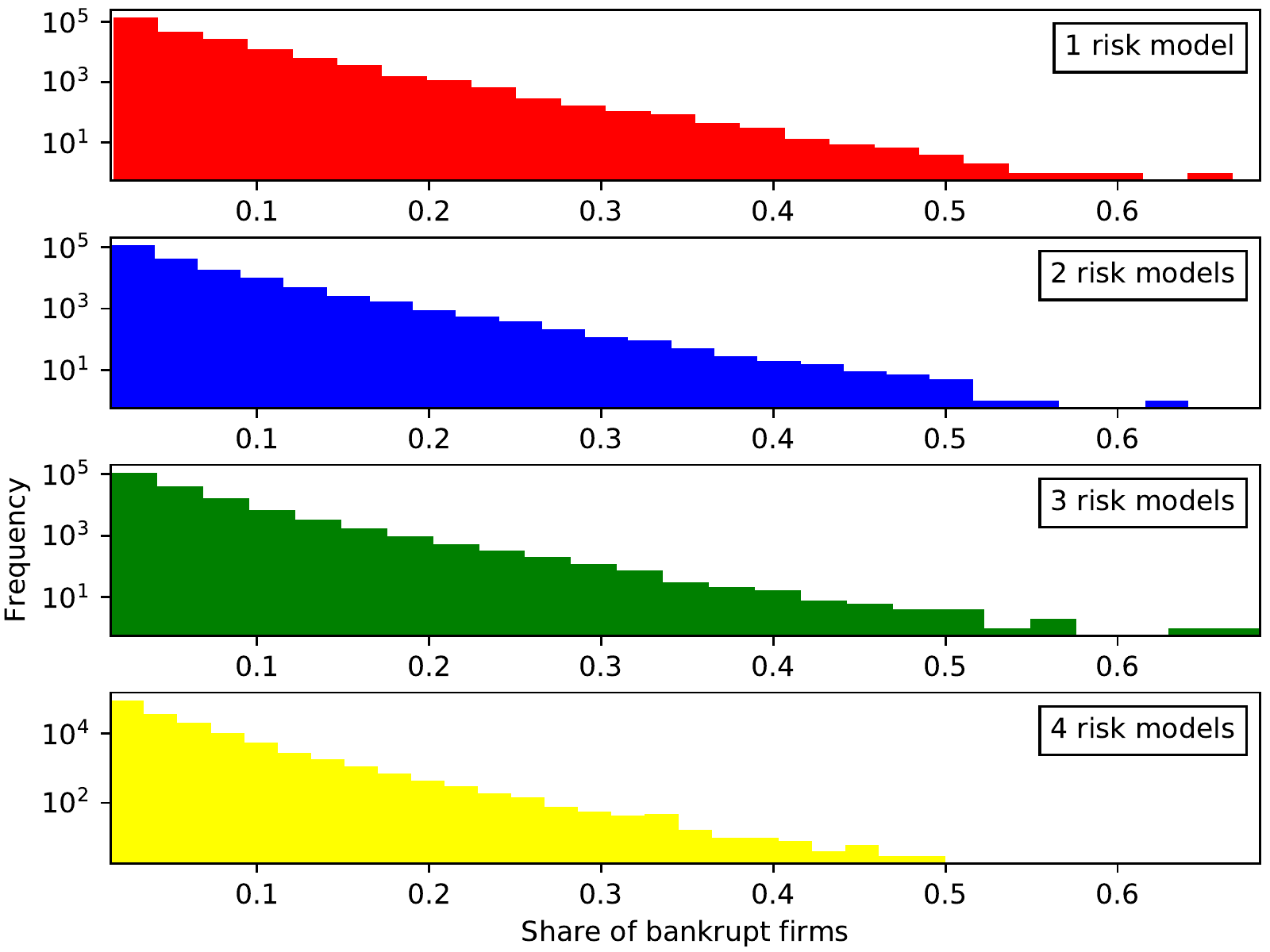}
  \caption{Density (histogram) of the total sizes of bankruptcy events (share of exiting firms $B$) in an ensemble of $400$ replications of simulations of $4,000$ time steps with margin of safety $\mu=1$. The y-axis is in log scale.}
  \label{fig:hist_b_mu1_re}
\end{figure}

\begin{figure}[tbhp]
  \centering
  \includegraphics[width=.8\textwidth]{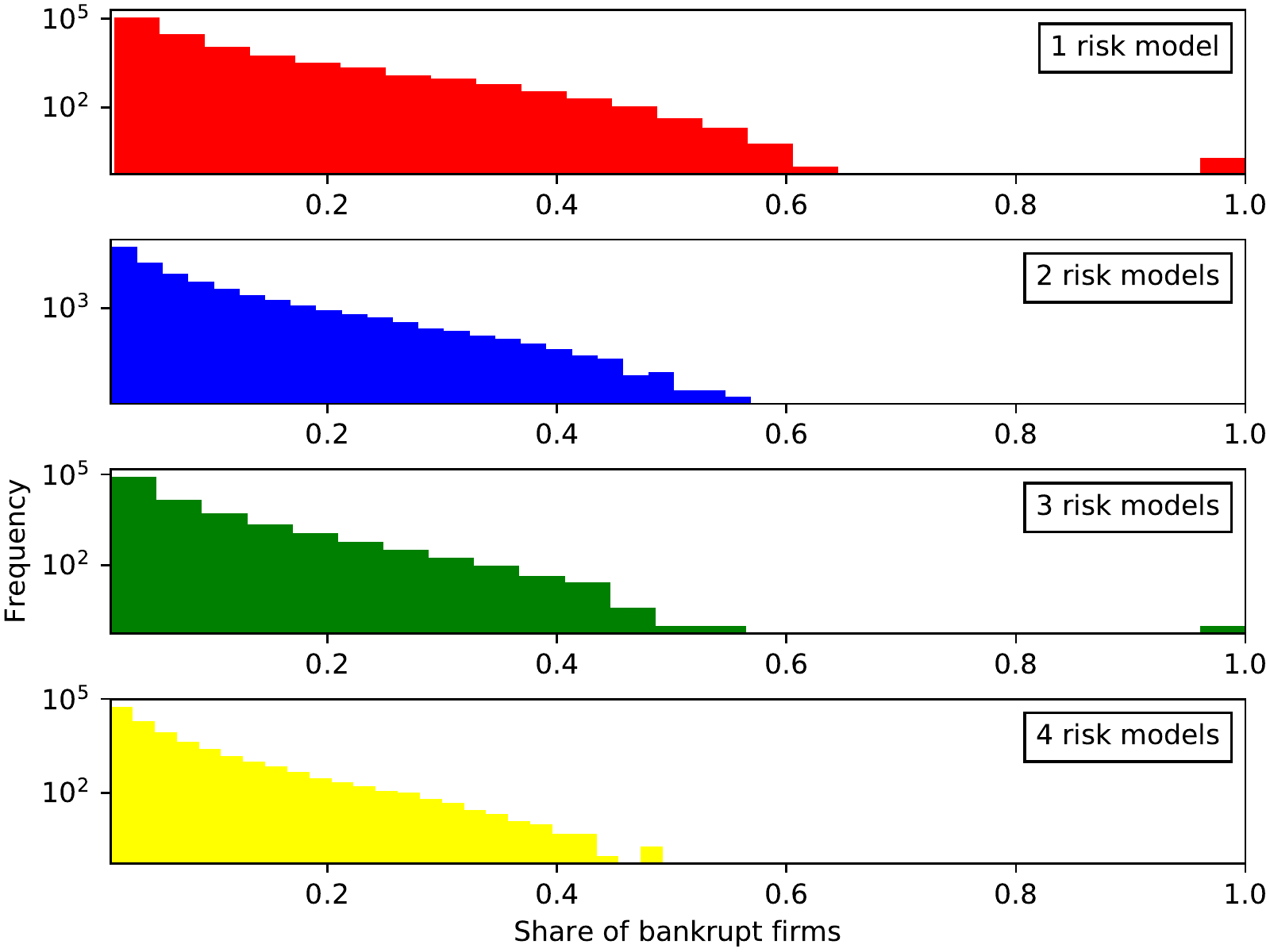}
  \caption{Density (histogram) of the total sizes of bankruptcy events (share of exiting firms $B$) in an ensemble of $400$ replications of simulations of $4,000$ time steps without reinsurance and with margin of safety $\mu=1$. The y-axis is in log scale.}
  \label{fig:hist_b_mu1_nore}
\end{figure}

We study the effect of the margin of safety $\mu$ on the system and its interaction with effects of risk model homogeneity. This effect can be seen in Figures~\ref{fig:hist_b_mu1_re} and \ref{fig:hist_b_mu1_nore}, where the margin of safety is reduced to $\mu=1$ in comparison to the standard case of $\mu=2$ shown in Figures~\ref{fig:hist_b_mu2_re} and \ref{fig:hist_b_mu2_nore}. This exacerbates the effect of risk model homogeneity and the associated systemic risk substantially and makes especially (but not only) cases with low risk model diversity more volatile. 

%

 \begin{figure}[tbhp]
   \centering
   \includegraphics[width=.9\textwidth]{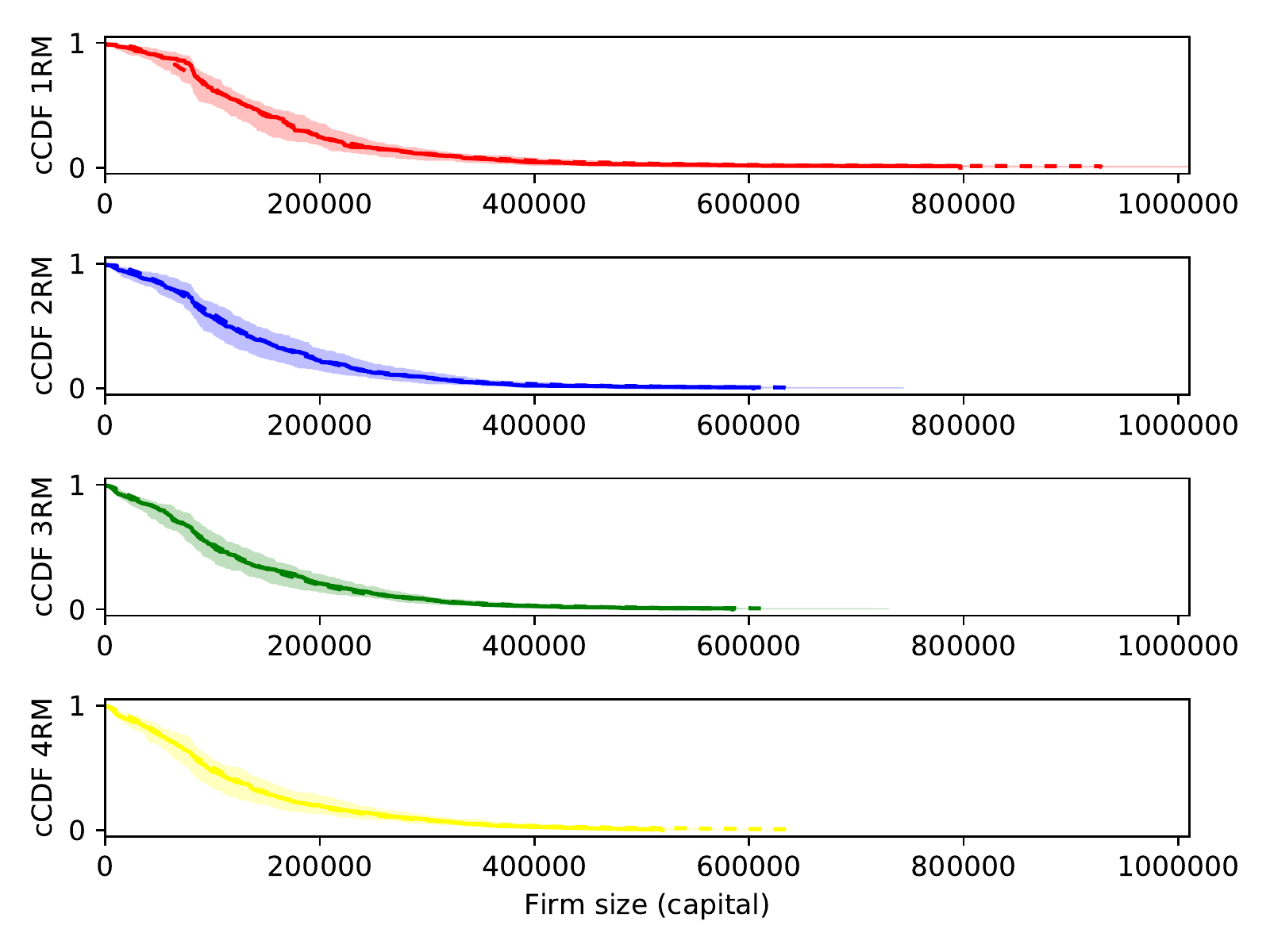}
   \caption{Ensemble of empirical complementary cumulative distribution functions (cCDFs) of the distribution of insurance firm sizes in terms of capital in an ensemble of $400$ replications of simulations of $4,000$ time steps without reinsurance and with margin of safety $\mu=2$. The median is shown as solid line, the mean as dashed line, the interquartile range as shaded area. Mean, median, and interquartile range are with respect to the ensemble of cCDFs, i.e., evaluated in x-axis direction.}
   \label{fig:ccdf_i_mu2_nore}
 \end{figure}

 \begin{figure}[tbhp]
   \centering
   \includegraphics[width=.9\textwidth]{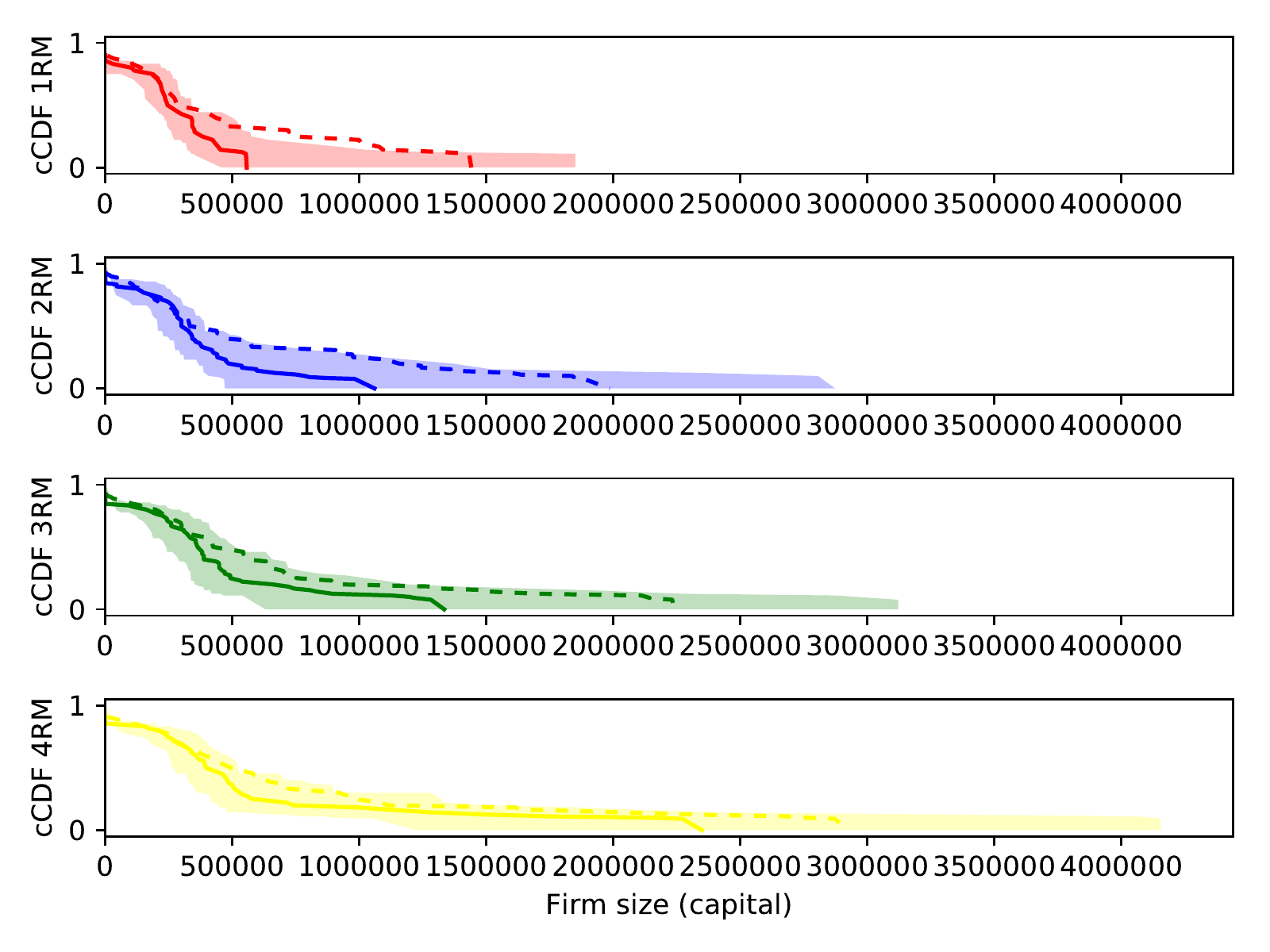}
   \caption{Ensemble of empirical complementary cumulative distribution functions (cCDFs) of the distribution of reinsurance firm sizes in terms of capital in an ensemble of $400$ replications of simulations of $4,000$ time steps with margin of safety $\mu=2$. The median is shown as solid line, the mean as dashed line, the interquartile range as shaded area. Mean, median, and interquartile range are with respect to the ensemble of cCDFs, i.e., evaluated in x-axis direction.}
   \label{fig:ccdf_r_mu2_re}
 \end{figure}

The emergence of asymmetric, long-tailed firm size distributions is preserved under a number of modifications including changing the margin of safety 
 and switching off reinsurance (see, Figure~\ref{fig:ccdf_i_mu2_nore}). The size distribution of reinsurance firms follows the same characteristics, albeit subject to more noise since the total number is smaller (see, Figure~\ref{fig:ccdf_r_mu2_re}).

Sensitivity analyses involving modifications to the interest rate $r$, the risk model inaccuracy parameter $\zeta$, the initial ratio of insurance to reinsurance firm capital $\overline{k_i}(0)/\overline{k_r}(0)$, and the runtime of contracts were conducted with smaller ensembles and shorter runtimes. The main result of the simulation were robust to the studied modifications. The smaller ensemble sizes do not support statements about differences in the quantitative results with reasonable confidence.

\end{document}